\documentclass[floats,preprint,aps,amssymb]{revtex4}
\usepackage{graphicx}
\usepackage{mathrsfs}
\DeclareGraphicsExtensions{.jpeg,.pdf,.eps,.gif,.tif,.sxd}
\DeclareGraphicsExtensions{.pdf}

\begin{document}
\title{Anomalous Lineshapes and Aging Effects
in  Two-Dimensional Correlation Spectroscopy }
\author{ Franti\v{s}ek \v{S}anda
$^{*}$
 and  Shaul Mukamel
$^{\dagger}$
}
\address{$^{*}$
Charles University, Faculty of Mathematics and Physics, Institute of Physics, Ke Karlovu 5,
Prague, 121 16 Czech Republic
\\  $^{\dagger}$
Department of Chemistry, University of California, Irvine, CA 92697-2025
}
\date{\today}
%\pacs{}
\begin{abstract}
\widetext
\bigskip

Multitime correlation functions provide useful probes for the ensembles of trajectories underlying the stochastic dynamics of complex systems.  These can be obtained by measuring their optical response to sequences of ultrashort optical pulse.  Using the continuous time random walk model for spectral diffusion, we analyze the signatures of anomalous relaxation in two-dimensional four wave mixing signals.  Different models which share the same two point joint probability distribution show markedly different lineshapes and may be distinguished.  Aging random walks corresponding to waiting time distributions with diverging first moment show dependence of 2D lineshapes on initial observation time, which persist for long times.

\vfill 
E-mail:\\
sanda@karlov.mff.cuni.cz \\
smukamel@uci.edu
\end{abstract}
\maketitle

\section{Introduction}
Simple relaxation theories break down when the relaxation is non-exponential and assumes, for example, a stretched-exponential or an algebraic form.  Such {\it anomalous relaxation} has been observed in numerous physical systems ranging from single molecules and quantum dot spectral diffusion in fluorescence blinking trajectories  \cite{xie2,amblard,verberk,nesbitt,bawendi,flomebom}, protein folding \cite{sabelko,oliveberg},
charge-carrier transport, geophysical processes, and in economics \cite{kirchner,iben,austin,mainardi,metzler1}.  Stochastic dynamics can be fully described by ensembles of trajectories of collective variables\cite{breuer}.  Statistical analysis of stochastic trajectories results in a hierarchy of multipoint correlation functions which carry increasing levels of information.  Two-point correlation functions provide the simplest measure of fluctuations and the most common evidence for anomalous relaxation.  They are the easiest to sample experimentally and to predict theoretically.  However they do not uniquely characterize the system.  Many models can be constructed that have the same two point correlations but very different higher order correlation functions.  Anomalous dynamics implies that many timescales are relevant.  These may represent various dynamical variables or metastable configurations in polymers or glassy systems \cite{frauenfelder,wolynes,granek}.  Treating all relevant variables explicitly is not always possible.  Some calculations only include directly accessible variables (such as the transition frequency in spectral diffusion) \cite{zwanzig} and use a master equation for their probability densities; all other variables are projected out and represented through memory functions.  The long time memories characteristic of anomalous relaxation are not compatible with the ordinary Markovian approximation which assumes  fast memory loss.  The master equations derived in this case \cite{kenkre,metzler} are thus limited to two point correlation functions and do not carry enough information to describe the multipoint correlation and response functions \cite{sanda4,nonequivalence}.

Several practical strategies may be employed towards the simulation of multipoint correlation functions.  One option is to use Markovian master equations with a large number of collective variables.  Another possibility is to assume harmonic (Gaussian) processes which are exactly solvable \cite{fox,leggett}.  All information is then contained in the spectral density, which may be tailored to give long-tailed correlations \cite{xie,granek}.  A different class of solvable models are continuous time random walks (CTRW) \cite{montroll,weiss}, which assume the erasure of all memory (renewal) when the relevant dynamical variables are changed (jumps).  They portray the dynamics as a generalized random walk with a distributed waiting time or length for stochastic jumps between various states.  Memory enters this model solely through the time $t$ elapsed from last renewal time.  Anomalous behavior is observed when the waiting time distribution function (WTDF)  $\psi(t)$ for the next jump has long tails.  We have recently proposed that lineshapes in coherent multidimensional optical spectroscopy may be used to probe anomalous multipoint correlation functions \cite{sanda3}.  Algebraic singularities at transition frequencies and power-law cross-peak dynamics were predicted in the two-dimensional optical response of a two-level chromophore to three laser pulses whose frequency undergoes a stochastic two state jump continuous time random walk with a power-law waiting time density function  $\psi(t)\sim t^{-\alpha-1}$.
In this paper we present more detailed simulations for this model and further demonstrate how it may be used to probe aging effects in systems that never equilibrate.  Frequency-domain signals such as linear absorption are ill defined in aging systems since they depend on the measurement time window.   2DCS is a time-domain technique that uses ultrashort pulses.  Such signals should provide unambiguous signatures for aging, since all delay times are fully controlled.  Different models may be distinguished by higher order nonlinear techniques.

We shall focus on two classes of WTDF which lead to anomalous spectral
lineshapes. We assume asymptotic algebraic decay $\psi (t)\sim t^{-\alpha
-1} $ which shows significant deviations from normal relaxation for $0<\alpha <2$
\cite{shlesinger,bouchard,klafter,shlesinger1}. For $1<\alpha <2$
 stationary ensembles may be described by a proper choice of initial
condition, which implies a special WTDF for the first jump $\psi ^{\prime
}(t)$ which represents how the system was prepared. The anomalous multipoint
correlations observed in fluorescence traces of conformation dynamics of
flavin proteins \cite{xie} showed symmetries due to microscopic
reversibility typical for stationary processes.

For $0<\alpha <1$, stationary ensembles can not be constructed. System
properties necessarily depend on the time elapsed from the initial preparation
even when it is very long. This phenomena is known as aging. Such random walks show
fractal behavior related to Levy stable distributions, which generalize the
Gaussian distributions of ordinary diffusion \cite{klafter}. This case is
fundamentally more complicated than $1<\alpha <2
$: such random walks are nonergodic \cite{bel},
their time and ensemble averages may differ \cite{margolin}, and
special sample preparation for each run of the experiment is needed.
Signatures of aging were observed in fluorescence blinking of single CdSe
quantum dots, with $\alpha \approx 0.5$ \cite{brookmann,nesbitt,bawendi}.
This is in agreement with the Sparre-Andersen theorem \cite{andersen,andersen1,andersen2}
which states that the first passage time of
random walk with any symmetric distribution of jump lengths (including Levy
flights) has a universal asymptotic $\sim t^{-3/2}$ decay. The origin of these
long tailed WTDF is not fully understood.

Environment dynamics affects  spectral lineshapes through modulations
of the transition frequencies. However, extracting the fluctuation
timescales from absorption lineshapes is not always possible and may require
additional assumptions and the introduction of specific models.
Nonlinear spectroscopies can distinguish between nonequivalent
dynamical models whose linear response is identical. In two-Dimensional
Correlation Spectroscopy (2DCS)\cite{mukamel,tanimura-mukamel,jonas} the
system is subjected to three femtosecond laser pulses (Fig 1). The first
pulse creates a coherence between the ground state and an excited state. Time
evolution (free induction decay) during the first interval $t_{1}$ is
related to the absorption lineshape by a Fourier transform. The second pulse
erases the coherence, bringing the molecule to the ground or an excited state
population. The transition frequency continues to change by interaction with
the environment during the second interval $t_{2}$. Finally a coherence is
again created by the third pulse and detected during the third interval $t_{3}$.
The various pathways for the density matrix  of a
two level chromophore  in Liouville space are shown in Fig 2. Correlations of the lineshapes
during the first and the third interval provide information on environment
dynamics during the intervening interval $t_{2}$. This supplements and greatly
expands the information obtained from linear techniques.

2DCS can monitor dynamical processes at the femtosecond timescale; analogous
2D NMR techniques are commonly used to study much slower (ms) processes \cite{ernst}.
Two-dimensional infrared lineshapes have been used to probe is the
structure of peptides \cite{hochstrasser}, the picosecond hydrogen bonding
dynamics by observing coherence transfer in molecular vibrations for phenol
in benzene \cite{fayer} and for acetonitrile in methanol \cite{kim}. In the
visible, 2DCS techniques have been used to study exciton transfer in
photosynthetic antennae \cite{fleming}.

Simulations of 2DCS signals usually employs either Markovian or Gaussian
models for spectral fluctuations. The response functions for Markovian
fluctuations may be obtained by the Green's function solution of the
Stochastic Liouville equations \cite{kubo,tanimura,anderson}, which combine
a Markovian master equation for jump dynamics with the Liouville equation
for coherent evolution. The response functions of a multilevel chromophore
linearly coupled to harmonic bath (Gaussian fluctuations) may be obtained by
the second order cumulant expansion using the Wick theorem. All higher
response functions may then be factorized into products of two point
quantities.

In the present work we
extend our earlier work \cite{sanda3} to study signatures of aging in 2D lineshapes.
In section II we build a general CTRW multistate jump model, and explain the condition of
microscopic reversibility. The theory of
2D lineshapes is presented in section III. In section IV we discuss various
parameter regimes of anomalous two state jump lineshapes. In
section V we study aging effects in 2CDS spectroscopy and
compare two approaches to describe aging: CTRW and time-dependent
markovian master equation. The two models have same
evolution of particle densities. The differences in 2DCS lineshape
thus reflect the role of the underlying trajectory picture, i.e. unravelling of the master equation \cite{breuer}, in multipoint probes.

\section{Two state CTRW jump model; stationarity,
microscopic reversibility, and aging}

In this section we briefly review the anomalous relaxation model used in \cite{sanda3}.
The multistate jump CTRW model is defined by a matrix $\hat{\Psi}(t)$ whose $ij$
element is the waiting time probability density function (WTDF) for
stochastic jumps from state $j$ to state $i$. $t$ is the time from the last jump
where all memory is erased. The matrix is normalized as
$\sum_{i}\int_{0}^{\infty }[\Psi]_{ij}(t)dt=1$.

In the simplest two state jump (TSJ) model
\cite{kubo,anderson,barkai,shushin1} bath has two states ($a$ and $b$). We represent connection
of density of renewals at various times in the $a,b$ space by the matrix:
\begin{equation}
\hat{\Psi}(t)=\left(
\begin{array}{cc}
0 & \psi (t) \\
\psi (t) & 0
\end{array}
\right)   \label{Psi}
\end{equation}
The survival probability $\phi _{i}(t)$ (that no jump had occurred from state $i$ for time $t$)
defines the diagonal matrix of survival probabilities
$\hat{\Phi}(t)$. It is connected to the waiting time density function $\psi
(t)$ by $\phi _{j}(t)=\int_{t}^{\infty }\sum_{i}\left[ \Psi \right]
_{ij}(t^{\prime })dt^{\prime }$. The survival probability matrix thus
connects the last renewal with final time
\begin{equation}
\hat{\Phi}(t)=\left(
\begin{array}{cc}
\phi (t) & 0 \\
0 & \phi (t)
\end{array}
\right) .  \label{Phi}
\end{equation}
The random walk is observed starting at time 0. The WTDF of the first jump
$\psi ^{\prime }(t)$ may differ from $\psi (t)$ since it depends on how the
system was prepared before $t=0$. Similarly, the $\hat{\Phi}^{\prime }(t)$
matrix represents the survival probability for the first jump.

For a stationary process, the density of jumps to state $i$, $\eta_i $ is connected to the
total density to be in state $i$, $\rho_i$ through
\[
\rho _{i}=\rho _{i}(t)=\int_{0}^{\infty }\eta _{i}(t-t^{\prime })
\phi_{i}(t^{\prime })dt^{\prime }=
\]
\[
=\eta _{i}\int_{0}^{\infty }\phi _{i}(t)dt=\eta_{i}
\int_{0}^{\infty }t\sum_{j}[\Psi ]_{ji}(t)dt=\eta _{i}\kappa _{1;i}
\]
where we have used the fact that all densities $\rho,\eta$ are time independent for stationary process and
$\kappa _{1;i}\equiv \int_{0}^{\infty }t\sum_{j}[\Psi ]_{ji}(t)dt$ is the
mean waiting time in the i-th state. It then follows that all $\kappa _{1;i}$ must
be finite. The rate for the  $j\rightarrow i$ jump is
\[
\eta _{j}\int_{0}^{\infty }[\Psi ]_{ij}(t)dt=\frac{\int_{0}^{\infty }[\Psi
]_{ij}(t)dt}{\kappa _{1;j}}\rho _{j}
\]
We can now define the rate coefficients 
\begin{eqnarray}
A_{ij}&\equiv& \frac{\int_{0}^{\infty }[\Psi]_{ij}(t)dt}{\kappa _{1;j}}; \quad \textrm{for} \quad  i\neq j
\nonumber \\
 A_{ii}&=&-\sum_{j; j\neq i} A_{ji}
\label{rates}
\end{eqnarray}
The stationary density $\rho _{j}$ is thus obtained by  the solution of the balance equation
\[
\sum_j A_{ij}\rho_j=0
\]

Using the same arguments, the WTDF for the first jump is
\begin{equation}
\lbrack \Psi^{\prime }]_{ij}(t)=\frac{1}{\rho _{j}}\int_{0}^{\infty }[
\Psi]_{ij}(t+t^{\prime })\eta _{j}(-t^{\prime })dt^{\prime }=\frac{
\int_{t}^{\infty }[\hat{\Psi}]_{ij}(t^{\prime })dt^{\prime }}{\kappa _{1;i}}
\label{first_jump}
\end{equation}

The stationary condition (Eq.(\ref{first_jump}) is closely related to
microscopic reversibility. CTRW is reversible if a trajectory $i_{1}$, $i_{2}
$, \dots $i_{n}$ with waiting times $\xi _{1}$,$\xi _{2}$, \dots , $\xi _{n}$
(last time is survival) is equally probable its reverse $i_{n}$, \dots ,
$i_{1}$ with waiting times $\xi _{n}$, \dots , $\xi _{1}$. We thus require
\[
\phi _{i_{n}}(\xi _{n})[\hat{\Psi}]_{i_{n}i_{n-1}}(\xi _{n-1})\ldots \lbrack
\hat{\Psi}]_{i_{3}i_{2}}(\xi _{2})[\hat{\Psi}^{\prime }]_{i_{2}i_{1}}(\xi _{1})\rho _{i_{1}}=\quad \quad \quad \quad \quad
\]
\begin{equation}
\quad \quad \quad \quad \quad =[\hat{\Psi}^{\prime }]_{i_{n-1}i_{n}}(\xi _{n})
[\hat{\Psi}]_{i_{n-2}i_{n-1}}(\xi _{n-1})\ldots \lbrack
\hat{\Psi}]_{i_{1}i_{2}}(\xi _{2})\phi _{i_{1}}(\xi _{1})\rho _{i_{n}}
\label{direct=reverse}
\end{equation}
for all paths (sequences and waiting times). Eq. (\ref{direct=reverse}) can
only be satisfied provided (i) the time profile of WTDF is independent of
jump direction $[\Psi ]_{ij}(t)=T_{ij}\psi _{j}(t)$ \cite{qian,note3}, (ii)
the rate coefficients for jump $A_{ij}=T_{ij}/\kappa _{1;j}$ ($i\neq j$) must satisfy detailed
balance \cite{schnakenberg}
\[
\frac{T_{ij}\rho _{j}}{\kappa _{1;j}}=\frac{T_{ji}\rho _{i}}{\kappa _{1;i}}
\]
and (iii) the probability of the  first jump and the last survival are related
through $\psi _{i}^{\prime }(t)=\phi _{i}(t)/\kappa _{1;i}$ which recovers Eq.(\ref{first_jump}).
Eq. (\ref{first_jump}) thus expresses microscopic
reversibility of a stationary ensemble: the survival probability coincides with the
probability for the first jump backward.

For the symmetric TSJ considered here (Eq. (\ref{Psi})) we simply have
\begin{equation}
\psi ^{\prime }(t)=\frac{\int_{t}^{\infty }\psi (t^{\prime })dt^{\prime }}{
\kappa _{1}}=\frac{\phi (t)}{\kappa _{1}}  \label{stat}
\end{equation}
and symmetric densities $\rho _{a}=\rho _{b}=1/2$.

For $0<\alpha <1$,  $\kappa _{1}$ diverges, and it is impossible to construct a stationary ensemble. 
Asymptotically the jump rate decreases to $1/\kappa _{1}$ which is
$0$ in this case \cite{aquino,barbi,shlesinger1}. This scenario applies for
arbitrary initial conditions.
Many properties now depend on the initial observation time (aging). The
normal diffusion constant for a Brownian particle moving on a lattice scales
as $\sim 1/\kappa _{1}$ and its variance grows linearly with time
(Einstein relation) $\langle \Delta x^{2}\rangle \sim t/\kappa _{1}$. When
$\kappa _{1}$ diverges, the particle loses its mobility at long times, and
its variance $\langle \Delta x^{2}\rangle $ growths grows sublinearly
$\sim t^{\alpha }$ (anomalous diffusion). Another remarkable point
is that the random walker survives at initial position for long times and
ergodicity is broken. As a corollary, time averages obtained in single
molecule measurements may be different from ensemble averages \cite{margolin}.

The simplest way to describe aging is by assuming that all random walks
start by a jump made at some time $t_{0}$ before the first laser pulse. The
common choice $\psi ^{\prime }(t)=\psi (t)$ implies 
$t_{0}=0$. The dependence on the initial observation time requires a $t_{0}$-dependent WTDF
$\psi ^{\prime }(t;t_{0})$. The consistent choice of $\psi ^{\prime }(t;t_{0})
$ will be discussed in section IV.

\section{Spectral diffusion in 2DCS signals}

We consider a two-level chromophore with a ground $|g\rangle $ and an
excited state $|e\rangle $, transition frequency $\Omega _{eg}$, and dipole
moment $\mu _{eg}$ subjected to three short laser pulses with an electric
field $E(t)$, and described by the Hamiltonian
\begin{equation}
H_{S}=|e\rangle \left[ \Omega _{eg}+\delta \Omega _{eg}(t)\right] \langle
e|-E(t)\mu _{eg}\left[ |g\rangle \langle e|+|e\rangle \langle g|\right]
\label{system}
\end{equation}
$\delta \Omega _{eg}(t)$ are stochastic frequency fluctuations caused by
interaction with the environment and described by the CTRW dynamics.
Observable quantities are obtained by averaging over all possible stochastic
paths of $\delta \Omega _{eg}(t)$ \cite{kampen}.

We associate the frequency fluctuations with  different bath states $i$,
each inducing a transition frequency shift $\delta \Omega _{i}$. In TSJ the transition frequency assumes the value $\delta \Omega _{eg}=\Omega
_{0}$ (state $a$) and $-\Omega _{0}$  (state $b$).

The response of our two level chromophore to three optical pulses is
described by the third order response functions. The various contributions
to the response function, known as Liouville space pathways (Fig 2), are
labelled $\nu $ \cite{principles}. During the intervals $t_{j}\equiv \tau
_{j}-\tau _{j-1}$, between successive laser interactions the system's
density matrix is in a given state $|\nu ^{(j)}\rangle =$ $|ee\rangle
,|gg\rangle ,|eg\rangle $, or $|ge\rangle $ with corresponding frequencies
$\Omega _{\nu }^{(j)}= $ $0,0,\Omega _{eg}$ and $-\Omega _{eg}$ respectively. 
The latter are modulated by the state of the bath.  The Liouville operator
describing the evolution in the bath state $|eg\rangle $; $\dot{\rho}_{eg}=\hat{L}_{eg} \rho_{eg}$ is given by the
following matrix in the $a,b$ space
\begin{equation}
\hat{L}_{eg}=\left(
\begin{array}{cc}
-i\Omega _{0} & 0 \\
0 & i\Omega _{0}
\end{array}
\right) ,  \label{repfreq}
\end{equation}
where $\hat{L}_{ge}=-\hat{L}_{eg}$, and $\hat{L}_{ee}=\hat{L}_{gg}=0$.

We next define the generating function $\rho _{\nu }$ by the
equation of motion.
\begin{equation}
\frac{d\rho _{\nu }}{dt}=-i\delta \Omega _{\nu }(t)\rho _{\nu }  \label{dfgf}
\end{equation}
with initial condition $\rho _{\nu }(0)=1$.
Here $\delta \Omega _{\nu }(t)=\delta \Omega ^{(j)}(t)$ for $t\in (\tau
_{j-1},\tau _{j})$. The third
order response function for the $\nu $'th pathway is then given by $R_{\nu
}^{(3)}(t_{3},t_{2},t_{1})\equiv \left\langle \rho _{\nu }\right\rangle $,
where $\langle \rangle $ implies averaging over the ensemble of bath paths.
Coherent signals are generated only in specific phase-matching directions.
Below we focus on the $\mathbf{k_{I}=-k_{1}+k_{2}+k_{3}}$ and
$\mathbf{k_{II}=k_{1}-k_{2}+k_{3}}$ directions. In the rotating wave
approximation these are represented by the four Liouville space pathways shown in Fig 2.

The $\mathbf{k_{I}}$ (photon echo) signal is \cite{principles}
%---------------------------------------------------------
\begin{equation}
{\mathscr{S}}_{I}(t_{3},t_{2},t_{1})=\left( \frac{i}{\hbar }\right) ^{3}\mu
_{eg}^{4}e^{-i\Omega _{eg}(t_{3}-t_{1})}\left[
R_{ii}(t_{3},t_{2},t_{1})+R_{iv}(t_{3},t_{2},t_{1})\right] ,  \label{k_1}
\end{equation}
%--------------------------------------------------------
and the $\mathbf{k_{II}}$ signal is
%--------------------------------------------------------
\begin{equation}
\mathscr{S}_{II}(t_{3},t_{2},t_{1})=\left( \frac{i}{\hbar }\right) ^{3}
\mu_{eg}^{4}e^{-i\Omega _{eg}(t_{1}+t_{3})}\left[
R_{i}(t_{3},t_{2},t_{1})+R_{iii}(t_{3},t_{2},t_{1})\right]   \label{k_2}
\end{equation}
%--------------------------------------------------------------
For stochastic models such as considered here the bath evolution and
equilibrium state are independent on the state of the system $\rho_{ee}$ or $\rho_{gg}$ so that 
$R_{i}=R_{iii}$, $R_{ii}=R_{iv}$.

The third order correlation function for the $\nu $'th pathway may be
obtained by solving Eq. (\ref{dfgf})
%-----------------------------------------------------------------
\begin{equation}
R_{\nu }^{(3)}(t_{3},t_{2},t_{1})\equiv \theta (t_{3})\theta (t_{2})\theta
(t_{1})\left\langle \exp \left[ {-i\int_{\tau _{2}}^{\tau _{3}}\delta
\Omega_{eg }(\tau _{3}^{\prime })d\tau _{3}^{\prime }}\right] \exp \left[ {
\mp i\int_{\tau _{0}}^{\tau _{1}}\delta \Omega _{eg}
(\tau_{1}^{\prime })d\tau _{1}^{\prime }}\right] \right\rangle   \label{mkorfor}
\end{equation}
%--------------------------------------------------------------------
where the upper sign represents $R_{i}=R_{iii}$ and the lower $R_{ii}=R_{iv}$.

%--------------------------------------------------------
The 2D signals are defined by  frequency-frequency $(\omega
_{3},\omega _{1})$ correlation plots for a fixed $t_{2}$.
%---------------------------------------------------------

\begin{equation}
S_{I}(\omega _{3},t_{2},\omega _{1})\equiv-Im\int \int \mathscr{S}_{I}(t_{3},t_{2},t_{1})
e^{i(\omega _{1}t_{1}+\omega _{3}t_{3})}dt_{1}dt_{3}
\end{equation}

\begin{equation}
S_{II}(\omega _{3},t_{2},\omega _{1})\equiv-Im\int \int \mathscr{S}_{II}(t_{3},t_{2},t_{1})
e^{i(\omega _{1}t_{1}+\omega _{3}t_{3})}dt_{1}dt_{3}
\label{mixdomain}
\end{equation}

We shall also display the following combination, which shows simpler
lineshapes with purely absorptive peaks \cite{tokmakoff,scheurer}.
%-------------------------------------------------------

\begin{equation}
S_A(\omega_3,t_2,\omega_1) \equiv S_I(\omega_3,t_2,-\omega_1)
+S_{II}(\omega_3,t_2,\omega_1)  \label{2DIRdef}
\end{equation}
%-------------------------------------------------------

The response is represented in $a,b$ space by a matrix $\hat{G}^{\nu }$
whose $jl$ element accounts for the contribution to $R_{\nu }^{(3)}$ from
paths with an initial bath state $l$ and final state $j$.
\begin{equation}
R_{\nu }^{(3)}(t_{3},t_{2},t_{1})=\sum_{jl}\left[ G^{\nu }\right]_{jl}
(t_{3},t_{2},t_{1})\left[ \rho _{\nu }\right] _{l}(t=0)  \label{defmult}
\end{equation}

For Markovian relaxation $[\Psi ]_{ij}(t)=T_{ij}e^{-t/\kappa_{1;j}}/\kappa_{1;j}$ each $\hat{G}^{\nu }$ may be factorized into a product
of three Green's functions representing the time evolution during the $t_{1}$
,$t_{2}$ and $t_{3}$ intervals whereby the density matrix is in the $\nu ^{(1)}
$, $\nu ^{(2)}$, and $\nu ^{(3)}$ states.
%---------------------------------------------------------
\begin{equation}
\hat{G}^{\nu }(t_{3},t_{2},t_{1})=\hat{G}^{\nu ^{(3)}}(t_{3})\hat{G}^{\nu
^{(2)}}(t_{2})\hat{G}^{\nu ^{(1)}}(t_{1})  \label{markovianfactorization}
\end{equation}
%---------------------------------------------------------
The Green's functions can be calculated by solving the stochastic Liouville
equations (SLE) \cite{tanimura}.
%-------------------------------------------------------------
\[
\frac{d\rho _{\nu }(t)}{dt}=\left( \hat{L}+ \hat{A}\right) \rho _{\nu }(t);
\]
%-------------------------------------------------------------
where $\hat{A}$ is the matrix of jump rate coefficients  (Eq. (\ref{rates})).
The SLE has recently been applied to describe vibrational 2D signals for
frequency fluctuations modulated by hydrogen bonding of phenol in benzene
\cite{sanda2}, conformation changes of peptides \cite{jansen} and infrared
lineshapes of water \cite{hayashi}.

The simulation of systems with long memory is much more complex. Various
types of reduced equations of motion for the CTRW dynamics have been
developed \cite{metzler,kenkre} for calculating the two-point correlation functions.
These, however, may not be extended to multipoint quantities required for
the description of 2DCS \cite{sanda1}, since the factorization, Eq.
(\ref{markovianfactorization}), does not hold for nonmarkovian relaxation.

We have recently \cite{sanda1} developed an algorithm for solving this model. 
 This is based on the successive recurrent construction of
a hierarchy of Green's functions. It relies on the renewal property
 computing the CTRW \cite{kampen}. Below we present an alternative, more
intuitive, derivation which is reminiscent of the Green's function method.

We need to maintain a bookkeeping of whether or not there was a jump during
each of the three time intervals $t_{1}$, $t_{2}$, $t_{3}$. For each of the
three intervals we must distinguish between two possibilities; either there was no jump or there
was a least one jump. $\hat{G}^{\nu }$ is thus given by a sum of $2^3=8$ terms
each representing one type of path in bath space.
%--------------------------------------------------------
\begin{equation}
\hat{G}^{\nu }(t_{3},t_{2},t_{1})=\sum_{m=1}^{8}\hat{G}_{m}^{\nu
}(t_{3},t_{2},t_{1})  \label{8terms}
\end{equation}
%-----------------------------------------------------
These terms are depicted in Fig 3, where the presence of any ($\geq 1$) jump
in a given time interval is represented by the trajectory touching the time
axis. 

$\hat{G}_{m}^{\nu }$ are conveniently recast in Laplace space. We
define (our notation is similar to \cite{shushin})
%---------------------------------------------------
\[
\hat{\tilde{\Psi}}(s-\hat{L})\equiv \int_{0}^{\infty }e^{-st}\hat{\Psi}
(t)\exp {\left( \hat{L}t\right) }dt
\]
%------------------------------------------------------
This implies for our TSJ model
%-------------------------------------------------------
\begin{equation}
\hat{\tilde{\Psi}}(s-\hat{L}_{eg})=\left(
\begin{array}{cc}
0 & \tilde{\psi}(s-i\Omega _{0}) \\
\tilde{\psi}(s+i\Omega _{0}) & 0
\end{array}
\right)   \label{mphi}
\end{equation}
%----------------------------------------------------------------
where $\tilde{\psi}(s)\equiv \int_{0}^{\infty }\psi (t)e^{-st}dt$ is the
Laplace transform of $\psi$. $\hat{\tilde{\Phi}}(s-\hat{L})$ for the
survival function is defined similarly
%------------------------------------------------------------
\begin{equation}
\hat{\tilde{\Phi}}(s-\hat{L}_{eg})=\left(
\begin{array}{cc}
\tilde{\phi}(s+i\Omega _{0}) & 0 \\
0 & \tilde{\phi}(s-i\Omega _{0})
\end{array}
\right)   \label{mpsi}
\end{equation}
%-----------------------------------------------------------
$\hat{G}_{m}^{\nu }$ is expressed as a matrix product of the propagators
through the intervals with any jump in the particular interval ( if the
trajectory touches the axis in Fig 3) , with additional factors for segments
connecting different intervals. These ensure that the bath
state does not change between the last jump in the earlier interval and the first jump in the
later interval. Both factors will be described below.

We first calculate the evolution for a fixed state of bath where
no jump occurs over several time intervals. Let us assume that the
state is fixed for time $t_{m}^{\prime }$ in the m-th interval,
till time $t_{l}^{\prime }$ in some subsequent l-th interval and
during all the intermediate intervals $t_{i} $, $l>i>m$. The
probability of this evolution is either $\hat{\Psi}$,
$\hat{\Psi}^{\prime }$,$\hat{\Phi}$, or $\hat{\Phi}^{\prime }$
depending on the path. The propagator connecting  the state
immediately after $t_{m}^{\prime } $ and after $t_{l}^{\prime }$
is given by
%-------------------------------------------------------------------
\begin{equation}
\hat{\Upsilon}(\Psi ,t_{l}^{\prime },t_{l-1},\ldots ,t_{m}^{\prime })=\hat{
\Psi}(t_{l}^{\prime }+t_{m}^{\prime }+\sum_{i=m+1}^{l-1}t_{i})\exp {\left(
\hat{L}^{(l)}t_{l}^{\prime }+\hat{L}^{(m)}t_{m}^{\prime
}+\sum_{i=m+1}^{l-1}\hat{L}^{(i)}t_{i}\right) }  \label{upsilon}
\end{equation}
%-------------------------------------------------------------------
where $\hat{L}^{(i)}=\pm \hat{L}_{eg},0$, depending on the state of the density matrix in the i-th interval. 
This contribution may appear in several ways. Either for the evolution
between the last jump in the m-th interval and the successive jump, first in
the l-th interval, or for the very first jump when the $t_{m}^{\prime }$
interval does not exist and $\Psi \rightarrow \Psi^{\prime }$. It also
appears for the survival from the very last jump when $t_{l}^{\prime }$
disappear and $\Psi \rightarrow \Phi $. Finally, when no jump occurs, then
$\Psi \rightarrow \Phi ^{\prime }$, $t_{l}^{\prime }$ is absent and 
$t_{m}^{\prime }=t_{1}$ .

 The second ingredient in our calculation
is the propagator through the k'th interval described by the
integral equation
%----------------------------------------------------------------
\begin{equation}
\hat{\Sigma} (\tau )=\int_{0}^{\tau}\hat{\Psi}(\tau -\tau ^{\prime
})\exp {\left[ -i\hat{L}^{(k)}(\tau -\tau ^{\prime })\right]
}\hat{\Sigma} (\tau ^{\prime })d\tau ^{\prime }
\label{propagation}
\end{equation}
%-----------------------------------------------------------------
with $\hat{\Sigma}(0)=\hat{1}$. The matrix $\Sigma$ connects the
arrival densities at two times within the same interval $t_{j}$.
By solving Eq. (\ref{propagation}) in Laplace space, we obtain the
following propagator through the k-th interval
%--------------------------------------------------------------------
\begin{equation}
\hat{\Sigma}(s_{k})\equiv \left[ 1-\hat{\tilde{\Psi}}(s_{k}-\hat{L}^{(k)})
\right] ^{-1}  \label{propagator}
\end{equation}
%-------------------------------------------------------------------
This contribution appears provided some jump had occurred in the k-th interval,
(i.e. the trajectory touches the axis in the k'th interval in Fig 3.) Eq.
(\ref{propagator}) can be interpreted as a summation of a geometric series
for paths with 1,2, \dots jumps in Laplace space, where time convolutions
become simple multiplications.

All of these factors should be convoluted in time to generate the trajectory. For instance, the domain
of integration for the first contribution $G_{1}^{\nu }$ is shown in Fig 4:
\[
\hat{G}_{1}^{\nu }(t_{3},t_{2},t_{1})=\int_{0}^{t_{3}}d\xi
_{6}\int_{0}^{t_{3}-\xi _{6}}d\xi _{5}\int_{0}^{t_{2}}d\xi
_{4}\int_{0}^{t_{2}-\xi _{4}}d\xi _{3}\int_{0}^{t_{1}}d\xi
_{2}\int_{0}^{t_{1}-\xi _{2}}d\xi _{1}
\]
\begin{equation}
\times \hat{\Upsilon}(\Phi ,\xi _{6})\hat{\Sigma}(t_{3}-\xi _{6}-\xi _{5})\hat{
\Upsilon}(\Psi ,\xi _{5},\xi _{4})\hat{\Sigma}(t_{2}-\xi _{4}-\xi _{3})\hat{
\Upsilon}(\Psi ,\xi _{3},\xi _{2})\hat{\Sigma} (t_{1}-\xi _{2}-\xi _{1})\hat{
\Upsilon}(\Psi ^{\prime },\xi _{1})  \label{allconvolutions}
\end{equation}
This results in a simple product in Laplace space
\begin{equation}
\hat{\tilde{G}}_{m}^{\nu
}(s_{3},s_{2},s_{1})=\hat{\tilde{\Upsilon}}(\Phi
,s_{3})\hat{\tilde{\Sigma}}(s_{3})\hat{\tilde{\Upsilon}}(\Psi
,s_{3},s_{2})
\hat{\tilde{\Sigma}}(s_{2})\hat{\tilde{\Upsilon}}(\Psi
,s_{2},s_{1})\hat{\tilde{
\Sigma}}(s_{1})\hat{\tilde{\Upsilon}}(\Psi ^{\prime },s_{1})
\label{res}
\end{equation}

We have already calculated Laplace domain $\tilde{\Sigma}$ (Eq.
(\ref{propagator})), $\Upsilon$ can be easily transformed as well,
leading to equivalent results to those reported Appendix C of Ref.\cite{sanda1}.
Eq. (\ref{res}) is finally expanded in terms of the matrices
$\hat{\Phi}$, $\hat{\Psi}$ and the complete expressions agrees
with Appendix D of Ref.\cite{sanda1}, where was obtained in
a different way.

Since the response functions (Eq. (\ref{mkorfor})) are causal, the 2D
lineshapes (Eq. \ref{mixdomain}) may be obtained by analytical continuation
of $s_{1},s_{3}$ (the Laplace variable conjugate to $t_{1}$ and $t_{3}$).
 The $t_{2}$ variable is obtained by reverse Laplace transform using
Bromwich integral
\[
S_{I}(\omega _{3},t_{2},-\omega _{1})=\frac{\mu _{eg}^{4}}{\pi \hbar ^{3}}
Im\int_{-i\infty }^{i\infty }ds_{2}e^{s_{2}t_{2}}\tilde{R}_{ii}\left(
s_{3}=-i(\omega _{3}-\Omega _{eg}),s_{2},s_{1}=i(\omega _{1}-\Omega
_{eg})\right)
\]

\[
S_{II}(\omega_3,t_2,\omega_1)= \frac{ \mu_{eg}^4}{\pi\hbar^3} Im
\int_{-i\infty}^{i\infty}ds_2 e^{s_2t_2} \tilde{R}_i(s_3=-i\left(\omega_3-
\Omega_{eg}),s_2,s_1=-i(\omega_1-\Omega_{eg})\right)
\]

For $t_{2}=0$ these integrals may be calculated analytically. The
resulting two-interval functions may be alternatively obtained by
directly building the two interval ($t_{3}$,$t_{1}$) response function.

\section{Lineshapes for stationary anomalous random walks}

Microscopic reversibility in stationary ensembles implies that
$\mathscr{S}_{I}(t_{3},t_{2},t_{1})=-\mathscr{S}_{I}^{\ast }(t_{1},t_{2},t_{3})$
which in
the frequency domain gives
\begin{equation}
S_{I}(\omega _{3},t_{2},-\omega _{1})=S_{I}(\omega _{1},t_{2},-\omega _{3})
\label{s1sym}
\end{equation}
Similarly $\mathscr{S}_{II}(t_{3},t_{2},t_{1})=\mathscr{S}_{II}(t_{1},t_{2},t_{3})$
which implies
\begin{equation}
S_{II}(\omega _{3},t_{2},\omega _{1})=S_{II}(\omega _{1},t_{2},\omega _{3}).
\label{s2sym}
\end{equation}

Combining Eqs (\ref{s1sym}) and (\ref{s2sym}) with Eq.
(\ref{2DIRdef}) we obtain the following symmetry of the lineshape
\begin{equation}
S_{A}(\omega _{3},t_{2},\omega _{1})=S_{A}(\omega _{1},t_{2},\omega _{3})
\label{sym_diagonal}
\end{equation}
Thus $S_{I}$, $S_{II}$, and $S_{A}$ are symmetric to the interchange of $\omega_1$ and $\omega_3$.

When during the  $t_{3}$ interval the bath has lost its memory of its
state during $t_{1}$ (e.g. normal relaxation with $t_{2}\rightarrow \infty $),
the response functions may be factorized as
%--------------------------------------------------------------
\begin{equation}
{\mathscr S}_{I}(t_{3},t_{2},t_{1})=2(i/\hbar )K(t_{3})K^{\ast }(t_{1})  \label{s1as}
\end{equation}
%-------------------------------------------------------------
and
%-------------------------------------------------------------
\begin{equation}{\mathscr S}_{II}(t_{3},t_{2},t_{1})=2(i/\hbar )K(t_{3})K(t_{1})  \label{s2as}
\end{equation}
%-------------------------------------------------------------
Here
%-------------------------------------------------------------
\[
K(t)\equiv (i/\hbar )\mu _{eg}^{2}e^{-i\Omega _{eg}t}\langle \exp
[-i\int_{0}^{t}\delta \Omega _{eg}(\tau )d\tau ]\rangle
\]
%--------------------------------------------------------------
is the linear response function for stationary ensembles. Its Fourier transform 
gives the absorption lineshape
\begin{equation}
W_{A}(\omega )\equiv Im\int_{0}^{\infty }K(t)\exp [i\omega t]dt
\label{linlsh}
\end{equation}
(The absorption of a nonstationary ensemble is not proportional to the Fourier transform of the
linear response function \cite{barkai}.)

Using Eqs. (\ref{s1as}), (\ref{s2as}), and (\ref{linlsh}), $S_A$ then reduces to
the product of the linear absorption lineshapes \cite{note1}
%----------------------------------------------------------
\begin{equation}
\hbar S_{A}(\omega _{3},t_{2}\rightarrow \infty ,\omega _{1})=4W_{A}(\omega_{1})
W_{A}(\omega _{3}).  \label{factorization}
\end{equation}
%----------------------------------------------------------
Algebraic memory decays will result in a slow convergence to this asymptotic
lineshape. In addition, as will be shown below, the spectra diverge at
certain frequencies where the factorization (Eq. (\ref{factorization})) does
not hold.

We shall consider a specific model of anomalous relaxation with the WTDF
\cite{sanda1,shlesinger}:
%------------------------------------------------------------
\begin{equation}
\tilde{\psi}(s)=\frac{1}{1+\kappa _{1}s/\left[ 1+(\kappa _{A}s)^{\alpha -1}\right] };\quad 1<\alpha <2;  \label{weak_anomalous}
\end{equation}
%--------------------------------------------------------------
$\kappa _{1}$ is the mean of $\psi (t)$, while $\kappa _{A}$ 
controls the long time algebraic tails $\psi _{W}(t)\sim \kappa _{A
}^{\alpha -1}\kappa _{1}/t^{\alpha +1}$.

Note that Eq. (\ref{stat}) may be conveniently represented in Laplace space
%------------------------------------------------------------
\[
\tilde{\psi}^{\prime }(s)=\frac{1-\tilde{\psi}(s)}{s\kappa _{1}}
\]
%-------------------------------------------------------------

We first consider the linear response obtained from the one-interval Green's
function
%------------------------------------------------------------
\[
K(t)=\sum_{jl}Q_{jl}(t)\rho (0)_{l}
\]
%------------------------------------------------------------
The kernel may be calculated by
%-----------------------------------------------------------
\[
\hat{Q}(t)=\hat{\Upsilon}(\Phi ^{\prime },t)+\int_{0}^{t}d\xi_{2}
\int_{0}^{t-\xi _{2}}d\xi _{1}\hat{\Upsilon}(\Phi ,\xi _{2})\hat{\Sigma}
(t_{1}-\xi _{2}-\xi _{1})\hat{\Upsilon}(\Psi^{\prime} ,\xi _{1}).
\]
%-----------------------------------------------------------
Transforming into the Laplace space domain yields \cite{sanda1,barkai,shushin}
%------------------------------------------------------------
\begin{equation}
\hat{\tilde{Q}}(s)=\hat{\tilde{\Phi}}^{\prime }(s-\hat{L})+\hat{\tilde{\Phi}}
(s-\hat{L})\left[ 1-\hat{\tilde{\Psi}}(s-\hat{L})\right] ^{-1}\hat{\tilde{
\Psi}}^{\prime }(s-\hat{L})  \label{linform}
\end{equation}
%-----------------------------------------------------------------

Combining Eqs. (\ref{linlsh}),(\ref{weak_anomalous}), and (\ref{linform}) we finally get
%----------------------------------------------------------------
\begin{eqnarray}  \label{wadshape}
W(\omega+\Omega_{eg})&=&\frac{2\Omega_0^2}{(\Omega_0^2-\omega^2)^2}
\nonumber \\
&\times& Re \frac{1}{\kappa_1+\kappa_{A}^{\alpha-1}\left[
(i\Omega_0-i\omega)^{\alpha-2} +(-i\omega-i\Omega_0)^{\alpha-2}\right]
+i(\omega-\Omega_0)^{-1}+i(\omega+\Omega_0)^{-1}}  \nonumber \\
\end{eqnarray}
%-----------------------------------------------------------------

In all plots we use dimensionless frequency units $(\omega _{j}-\Omega
_{eg})/\Omega _{0}$ by setting $\Omega _{eg}=0,\Omega _{0}=1.$ In Fig 5 we
display the absorption spectrum in the slow ($\kappa
_{1}\Omega _{0}>1$, top) and the fast ($\kappa _{1}\Omega _{0}<1$, bottom)
fluctuation limits. The lineshape has two peaks at $\omega =\pm 1$ and in
the fast fluctuation limit we obtain a finite central peak \cite{sanda1,barkai}.
The fraction of particles that remained at the initial
position is significant (not exponentially small) at all times. This
results in the divergence of peaks at $\omega =\pm 1$ 
%----------------------------------------------------------
\begin{equation}
W(\omega )\approx C|\Delta \omega |^{\alpha -2}
\end{equation}
%---------------------------------------------------------
with 
$C=\cos{\left[ \pi (1-\alpha /2)\right] }\kappa _{A}^{\alpha -1}/2$, and 
where the detuning is $\Delta \omega \equiv \omega -\Omega _{eg}-\Omega _{0}$ for $\omega =1$
and $\Delta \omega \equiv \omega -\Omega _{eg}+\Omega _{0}$ for $\omega =-1$
peak \cite{sanda1,barkai} .

The parameter $\alpha $ controls the peak singularity. For $\alpha
\rightarrow 2$ the divergence is cured and we approach the
Markovian lineshape. For fast fluctuations $\Omega _{0}\kappa
_{1}<<1$ the central peak grows, as $\kappa_1$ becomes shorter. This is reminiscent of the
motional narrowing for the Markovian case. However, the two divergent peaks
 still retain an anomalous lineshape. 

In Fig. 6A we display the $S_{I}(\omega_{3},
-\omega _{1})$, $S_{II}(\omega _{3},\omega _{1})$ and $S_{A}(\omega_{3},
\omega _{1})$ signals for slow fluctuations $\Omega _{0}\kappa _{1}>>1$
and $t_{2}=0$. Similar to the Markovian case \cite{sanda2},
all panels show two diagonal-peaks at $(\omega _{3},\omega _{1})=(1,1)$ and
$(-1,-1)$. However the peaks are nonlorentzian and divergent. $S_{I}$ and
$S_{II}$ diverge along the $\omega _{1}=\pm 1$, $\omega _{3}=\pm 1$ lines,
but much of this divergence is cancelled in $S_{A}$ which only diverges at
peaks (1,1), and (-1,-1).

We next examine the analytic structure of these divergencies for the $\omega_1=1$ and $\omega_3=1$ lines.
The slowest decay is connected with the survival function for the first jump
$\phi ^{\prime }(t)\sim t^{1-\alpha }$. $\hat{G}_{\nu }^{8}$ is thus the
most rapidly divergent term. The analysis of peak divergencies thus reduces
to the $\hat{G}_{\nu }^{8}$ contribution. We denote $\Delta \omega_{3}
\equiv \omega _{3}-\Omega _{eg}-\Omega _{0}$ and $\Delta \omega_{1}
\equiv \omega _{1}-\Omega _{eg}-\Omega _{0}$ and find
%------------------------------------------------
\[
S_{I8}(\omega _{3},t_{2}=0,-\omega _{1})=-\frac{\mu ^{4}}{\hbar ^{3}}Im
\frac{\tilde{\phi}^{\prime }(-i\Delta \omega _{3})-\tilde{\phi}^{\prime
}(i\Delta \omega _{1})}{\Delta \omega _{1}+\Delta \omega _{3}}
\]
%---------------------------------------------
\begin{equation}
S_{II8}(\omega _{3},t_{2}=0,\omega _{1})=\frac{\mu ^{4}}{\hbar ^{3}}Im\frac{
\tilde{\phi}^{\prime }(-i\Delta \omega _{3})-\tilde{\phi}^{\prime
}( -i\Delta \omega _{1})}{\Delta \omega _{1}-\Delta \omega _{3}}
\label{contrib}
\end{equation}
%------------------------------------------------
The lineshapes (Eqs.(39)) diverge along the lines
$\Delta \omega _{3}=0$ and $\Delta \omega _{1}=0$. The divergent
peak structure is  summarized in Table I. The left column corresponds to situation when $\Delta \omega _{3}$ is held fixed at a
small but nonzero value and  $\Delta \omega _{1}$ approaches the singular point
 0. Thus we consider $\Delta \omega _{1}<<\Delta \omega_{3}$
and $\tilde{\phi}^{\prime }(i\Delta \omega _{1})>>\tilde{\phi}^{\prime
}(-i\Delta \omega _{3})$. With the asymptotic expansion
%--------------------------------------------------------
\begin{equation}
\tilde{\phi}^{\prime }(s)\sim \kappa _{A}^{\alpha -1}s^{\alpha -2}
\label{expansion}
\end{equation}
%-----------------------------------------------------
we get the asymptotic form of divergent $S_{I}(\omega _{3},-\omega _{1})$, and $S_{II}(\omega _{3},\omega _{1})$ shown in the
Table I. In the right column we similarly approach the singular line at $\Delta \omega _{3}=0$.
We have verified these analytic asymptotic results numerically (not shown, it also qualitatively agrees with Fig 6A.

$S_{I}$ and $S_{II}$ have opposite signs, and their combination $S_{A}$ is
finite due to interference. The divergencies are only seen at the (1,1) and (-1,-1) peaks,
and not along the entire $\omega _{1}=\pm 1$ and $\omega _{3}=\pm 1$ lines,
since the $S_{I},S_{II}$ divergencies cancel. $S_{A}$ is finite, but
nondifferentiable along these lines.

We next examine more closely the variation along the $\Delta \omega _{1}=0$
axis.
\[
S_{A8}(\Delta \omega _{3},t_{2}=0,\Delta \omega _{1}=0)=\frac{-2\mu _{eg}^{4}}{\hbar ^{3}}
\frac{Im\tilde{\phi}^{\prime }(i\Delta \omega _{3})}{
\Delta \omega _{3}}
\]
The asymptotic expansion Eq.(\ref{expansion}) yields the analytical peak
structure at $\Delta \omega _{3}\approx 0$.
%-----------------------------------------------------------
\begin{equation}
S_{A}(\Delta \omega _{3},t_{2}=0,\Delta \omega _{1}=0)\approx B \Delta \omega _{3}^{\alpha -3}  \label{asympt_peak}
\end{equation}
\[
B=\frac{2\mu_{eg}^{4}}{\hbar ^{3}}\kappa _{A}^{\alpha -1}\sin {\left[ \pi
(2-\alpha )/2\right] }
\]
%-----------------------------------------------------------
The analytic structure of the (-1,-1) peaks is the same. This follows from
the assumed $[\hat{\Psi}]_{ab}=[\hat{\Psi}]_{ba}$ symmetry of TSJ model, which implies
$S_{\nu}(\omega_3+\Omega_{eg},t_2,\omega_1\mp\Omega_{eg})=S_{\nu}(-\omega_3+\Omega_{eg},t_2,-\omega_1\mp\Omega_{eg})$;
upper sign applies for $\nu=I$ lower for $\nu=II,A$.
Based on Fig 6a, the peaks are more localized with steeper contours for smaller
$\alpha $. In all cases we see a dip at (0,0).

The two peaks induced by $\hat{G}_{8}$ are universal and survive even for
the case of fast fluctuations $\Omega _{0}\kappa _{1}<<1$, as shown at Fig 6B. Rapid
changes during $t_{1}$ and $t_{3}$ induce a new peak at the average frequency
(0,0), (motional narrowing). The $S_{A}$ (0,0) peak is Lorentzian: The
star-like contours, best seen for $\alpha =1.2$ correspond to a product two
Lorentzians along $\omega _{1}$ and $\omega _{3}$. The $S_{I}$ and $S_{II}$
lineshapes are similar. Both may be described by a
statistical mixture of rapidly fluctuating particles responsible for the
central peak, with the static phase responsible for the divergent peaks at
the fundamental frequencies. Surprisingly, this picture is most pronounced
for small $\alpha =1.2$ where all peaks are well-separated. Increasing
$\alpha $ broadens the (-1,-1), and (1,1) peaks, making them interfere with
the central peak, and the lorentzian shape becomes less pronounced as $\alpha
\rightarrow 2$.

The variation of $S_{A}$ with $t_{2}$ in the slow fluctuation limit is
displayed in Fig 7. For $t_{2}$ longer than the mean waiting time $\kappa
_{1}$ fractions of trajectories have different frequencies in
the $t_{1}$ and $t_{3}$ intervals , as described by the $G_{6}$
contribution resulting in new cross peaks at (-1,1); (1,-1). Since we are in
the slow fluctuation limit the peaks are still well resolved. Both diagonal
and cross peak contours are elongated along the $\omega _{1,3}=\pm 1$
directions. Nevertheless the decay of the $G_{6}$ contribution
$t_{1,3}^{\alpha -3}$ (compared to the diverging $t_{1,3}^{\alpha -2}$ decay
of $G_{8}$ which is relevant for diagonal peaks) is integrable and thus the
cross peaks do not diverge. Another notable point is the
breakdown of Eq.(\ref{factorization}) at $\omega _{1,3}=\pm 1$; Memory
loss is not complete since the algebraic functions do not factorize.
At other frequencies the lineshapes approach this
limiting lineshapes  (Eq. (\ref{factorization})) algebraically as $t^{1-\alpha} $
\cite{sanda3}. These simulations illustrate the capacity of 2DCS to probe
anomalous relaxation during the $t_{2}$ interval.

\section{Nonstationary ensembles; Aging of 2D lineshapes}

In our earlier work \cite{sanda3} we considered 
nonstationary ensembles with $0<\alpha <1$ by assuming that the random walk is
started by a jump at the time origin, coinciding  with the first
laser pulse, so that response may be calculated by $\psi'=\psi$.
The lack of microscopic reversibility is reflected in violations
of the symmetry relations Eq. (\ref{sym_diagonal}). The higher
mobility during the (earlier) $t_{1}$ interval compared to $t_{3}$
resulted in broader peaks along the $\omega _{1}$ axis compared to
$\omega _{3}$.

Here we explore signatures of aging.
 We consider random walks, which  start by a
jump made at some time $t_{0}$ before the first laser pulse and examine how the nonlinear lineshapes vary with $t_0$.
The
response function then depends $t_{0}$ even for $t_{0}\rightarrow
\infty $. This is known as aging. All aging effects are fully described by calculating the
WTDF $\Psi ^{\prime }(t;t_{0})$ for the first jump which is now
$t_{0}$ dependent which must be consistent
with the CTRW dynamics during the $t_{0}$ period.

$\Psi ^{\prime}(t;t_{0})$ can be calculated along the lines of Eq.
(\ref{upsilon}) by omitting the coherence evolution $\hat{L}^{(0)}=0$
during $t_0$,
\[
\hat{\Psi}^{\prime}(t,t_{0})= \hat{\Psi}(t+t_{0})+ \int_0^{t_{0}} d \xi_2
\int_0^{t_{0}-\xi_2} d\xi_1 \hat{\Psi}(t+\xi_2)\hat{\Sigma}
(t_{0}-\xi_2-\xi_1)\hat{\Psi}(\xi_1)
\]

In Laplace space we find for our TSJ model
\begin{equation}
\tilde{\psi}^{\prime }(s;s_{0}) =\frac{[\tilde{\psi} (s_{0})-\tilde{\psi}
(s)]}{[1-\tilde{\psi}(s_{0})](s-s_{0})}
\label{delayed}
\end{equation} The 2D lineshapes may thus be calculated using the algorithm
presented in Section III. The $t_0$ dependence is obtained by
numerically inverting these Laplace domain formulas.

The long $t_{0}$ limit may be obtained by setting $s_{0}\rightarrow 0$. For
CTRW with finite $\kappa _{1}$ the denominator in Eq. (\ref{delayed}) is
\[
1-\tilde{\psi}(s_{0})\approx -s_{0}\frac{d\tilde{\psi}(s)}{ds}{\bigg|}_{s=0}
=\kappa _{1}s_{0}
\]
This reproduces the WTDF of the first jump for a stationary random walk
$\psi ^{\prime }(t;t_{0})=\phi (t)/\kappa _{1}$.

The lack of stationarity has some important consequences. As pointed in \cite{barkai}
frequency domain absorption measurement is no longer given by the Fourier transformed response function.
 Thus the absorption of an aging ensemble can not be calculated using Eq. (\ref{linform}).
Fortunately,  2DCS  works in the time domain, and the
measurement directly probes the response function. Thus the problems
discussed in \cite{barkai} do not apply for impulsive time-domain techniques such as 2DCS lineshape.

A more subtle point is that due to the lack of equilibration, the averaging over
consecutive pulse sequences may depend on the experimental data acquisition
repetition rate. Proper definition of the response function requires a
careful preparation $\psi ^{\prime }(t)$ before each pulse sequence.

We have calculated the response functions 
the variation of the lineshape with the preparation time
$t_{0}$ for the following model
\begin{equation}
\tilde{\psi}_N(s)=\frac{1}{1+(\kappa s)^{\alpha}}, \quad \alpha \in (0,1)
\label{mittag}
\end{equation}
This corresponds to a WTDF with algebraic tails $\psi
(t)\sim(\kappa/t)^{1+\alpha }$.

We took $\alpha =0.98$, which is close to the
Markovian case (Eq. (\ref{mittag}) for $\alpha =1$) in the fast
fluctuation limit $\kappa \Omega _{0}<<1$. This choice is
motivated by the simpler interpretation of the lineshapes; we expect
it to be closer to the Markovian case than the rather complex
$t_0=0$ shapes presented in \cite{sanda3}. The effect of $t_{0}$
could thus be better isolated. In addition, aging effects appear at arbitrarily long timescales (for suitable choice of parameters).  This
overcomes the difficulty with strongly anomalous ensembles, whose lineshapes cannot be obtained by repeated measurements on
the same sample, whose response function is changed between
two pulse sequences.

The top left panel of Fig. 8 ($t_{0}=0$) shows fast-fluctuation
Markovian contours and only tiny peaks at (1,1), and (-1,-1). 
No signatures of time irreversibility are seen since Eq.
(\ref{sym_diagonal}) is nearly satisfied. We next increase the
aging time $t_{0}$ as we move from the top left panel to bottom right
panel. The (1,1) and (-1,-1) peaks appear and
grow, while the central peaks slowly get weaker. This reflects
decrease of the jump rate with time. Some small deviation from
the symmetry relation Eq. (\ref{sym_diagonal}) can be noticed. The
process is nearly reversible on the $\Omega _{0}^{-1}$ timescale which
dominates the lineshapes.
A remarkable point is that the central (motional narrowing) peak
coexists with these static limit peaks. The anomalous process is
better viewed as a mixture of static and fluctuating particles,
rather than a homogeneous rate.

This clearly distinguishes our algorithm from calculations based
on time-dependent rate master equations, which do not allow
to properly describe memory effects in multipoint probes. To
support this statement we have constructed Markovian process
subjected to the same master equation, i.e. we require correct
prediction of total densities and subsequently apply them to
calculating response or multipoint correlation function based on
Markovian schemes. The trajectory picture of both approaches is
different \cite{twoapproaches}.

Consider a Markovian master equation whereby densities evolve in the
same way as the aging random walk for arbitrary initial densities,
i.e. it has the same Green's function $G(t)$ .
\begin{equation}
 \rho(t)=G(t)\rho(0)
 \label{ME}
\end{equation}
The master equation is constructed by differentiating Eq.(\ref{ME})
with respect to time
\begin{equation}
\frac{d \rho(t)}{dt}=A(t)\rho(t);\quad \quad   A(t)\equiv \frac{d
G(t)}{dt}G^{-1}(t). \label{TCLME}
\end{equation}
The transition matrix
$A$ of time-convolutionless master equation is thus uniquely defined.
The Green's function Eq.(\ref{ME}) is the solution of the
mater equation
\begin{equation}
G(t)=\exp_T{\int_0^{t} A(t')dt'} \label{solution}
\end{equation}
We consider a symmetric two state dynamics parametrized by
a single  function $\Lambda$
\[
A(t)=\left(
\begin{array}{cc}
-\Lambda(t) &\Lambda(t) \\
\Lambda(t) & -\Lambda(t)
\end{array}
\right)
\]
Eq.(\ref{solution}) can be solved after a simple algebra. This gives
\begin{equation}
G_{11}(t)-G_{10}(t)= \exp{[-2\int_0^t \Lambda(t')dt']} \label{GL}
\end{equation}
Inverting Eq.(\ref{GL}), the rates can be calculated once the Green's
function is known
\begin{equation}
\Lambda(t)=\frac{-\frac{d}{dt}\left[G_{11}(t)-G_{10}(t)\right]}{2[G_{11}(t)-G_{10}(t)]}
\label{defrates}
\end{equation}
We next adjust the Green's function to agree with those of  our
aging random walk. In Laplace space it reads
\[
G_{11}(s)-G_{10}(s)=\frac{\phi(s)}{1+\psi(s)}=\frac{1-\psi(s)}{s[1+\psi(s)]}
\]
For the model Eq. (\ref{mittag}),
\[
G_{11}(s)-G_{10}(s)=\frac{(\kappa s)^{\alpha}}{s[1+(\kappa
s)^{\alpha}]}
\]
 which may be also calculated directly in time domain as series
\begin{equation}
 G_{11}(t)-G_{10}(t)=\sum_{n=0}^{\infty}(-1)^n\frac{(t/\kappa)^{\alpha
 n}}{\Gamma(n\alpha+1)}
 \label{td}
\end{equation}
with the gamma function $\Gamma(y)\equiv \int_0^{\infty}x^{y+1} e^{-x}
dx$. The master equation is thus defined by combining Eqs.
(\ref{TCLME}), (\ref{defrates}), and (\ref{td}). The rates decay
asymptotically ($t\rightarrow \infty$) as $\Lambda(t)\approx
\alpha/(2t)$. Exponential WTDF's ($\alpha=1$) correspond to
constant rate $\Lambda=\kappa^{-1}$.

 Fig 9. shows the time dependent rate of the master equation for various $\alpha$.
 Aging effects (decreasing mobility with time) are reflected in the decreasing rates.
 Increasing $\alpha$ the decay is slower when approaching the markovian
limit ($\alpha=1$) and the rates change slowly for long
periods. This regime is particularly interesting because it may
provide sufficient time to measure the rate constant
by e.g. lineshape experiments and give clear meaning to our
arguments. (Diverging rates at very small times $t<<\kappa$ are
integrable and thus insignificant.)

We shall compare two types of stochastic processes
subjected to the same master equation, but with different
unravelling into trajectories \cite{breuer}. Aging lineshapes for  the CTRW model were already
presented at Fig. 8.
The second model is defined by Markovian prescription: 
The probability of jumps are independent of the
past trajectory. The stochastic Liouville equations and Green's
function technique may then be used to calculate the nonlinear response.
\[
R_{\nu}(t_3,t_2,t_1;t_0)=\theta (t_{3})\theta (t_{2})\theta
(t_{1}){\bigg \langle}  \exp_T \left[ {\int_{\tau _{2}}^{\tau _{3}}
\left[A(\tau_{3}^{\prime })+\hat{L}^{(3)} \right] d\tau
_{3}^{\prime }}\right] \exp_T \left[ {\int_{\tau _{1}}^{\tau _{2}}
A(\tau_{2}^{\prime })  d\tau _{2}^{\prime }}\right]
\]
\[\times
 \exp_T \left[ {  \int_{\tau _{0}}^{\tau _{1}}
\left[A(\tau_{1}^{\prime })\pm\hat{L}^{(1)} \right]d\tau
_{1}^{\prime }}\right] {\bigg \rangle}
\]

We are interested in the peak pattern, which is influenced by
fluctuations on the $\Omega_0^{-1}$ timescale. We consider a parameter
regime where the rate does not change significantly on this
timescale, and thus the peak pattern may be analyzed by a 
simple approximation of rates independent of $t_1$, and $t_3$ and
analyze aging of $t_2=0$ lineshapes
\[
R_{\nu}(t_3,0,t_1;t_0)=\theta (t_{3})\theta (t_{2})\theta
(t_{1})\left\langle \exp \left[\left(A(t_0)+\hat{L}^{(3)}
\right)t_3\right] \exp \left[\left(
 A(t_0)\pm\hat{L}^{(1)} \right)t_{1}\right] \right\rangle
\]
We then obtain
\begin{equation}
 S_{\beta}(\omega_3,0,\omega_1;t_0) = \frac{2\mu^4}{\hbar^3} Re
\frac{4\Lambda^2-\omega_3\omega_1\mp\Omega_{0}^2
-i2\Lambda(\omega_3+\omega_1)}{\left[\omega_1^2-\Omega_{0}^2+i2\omega_1\Lambda\right]\left[\omega_3^2-\Omega_{0}^2+i2\omega_3\Lambda\right]}
\label{ml}
\end{equation}

 where the upper (lower) sign is for $\beta=I$ ($\beta=II$)
and where $\Lambda \equiv \Lambda (t_0)$.

The aging Markovian  2D absorptive lineshapes are presented at Fig 10. The central peak is gradually broadened with
increasing time (and decreasing rates) and splits into
two peaks centered along diagonal at fundamental frequency. These peaks get narrower
for long $t_2$.

The significance of the different trajectory picture can be seen by comparing the
two lineshapes at Fig 8 and 10. We
 notice that the crossover to static lineshapes is somewhat faster at Fig 10. This
 is, however, less obvious feature, since it depends on chosen particular parametrization.
  The more significant feature, which distinguishes the two models
 is that the static peaks at fundamental frequencies and the fast
 motional narrowing central peak never coexist at Fig 10 in contrast to Fig 8.

 This may be explained as the direct signature of memory.
The CTRW model shows two populations static and fast fluctuating, i.e.
 particles are differentiated based on their histories.
In contrast all particles in the
 Markovian model have homogenous probabilities for the next jump.
 This lack of memory is reflected in the unique peak pattern
with no simultaneous static and fast fluctuating signatures in the
spectrum.

These two models are nonequivalent since they assign
different trajectory picture to the same density matrix, as is clearly seen from the higher order correlation functions and response.
The coexistence of both static and fast fluctuations in spectra
clearly reflects the additional information, beyond the two point
correlation functions.

The unravelling of master equations into trajectories is an important 
issue.  Two dimensional lineshapes which are sensitive to the 
trajectories should provide a direct test for the unravelling schemes \cite{breuer}.
Single molecule spectroscopy looks at the trajectories one at a time.
Multidimensional spectroscopy looks at the entire ensemble but unravels it by the manipulation of coherence.

 In summary, our simulations demonstrate that two-dimensional
correlation plots of signals obtained from the response of the
system to sequences of multiple laser pulses carry specific and
direct signatures of complex dynamics. Such techniques are
currently feasible in many spectral regimes, NMR, EPR, the infrared (vibrations, phonons) and
in the visible (electronic excitations).

\acknowledgements The support of the Ministry of Education, Youth and Sports
of the Czech Republic (project MSM 0021620835), GA\v{C}R (Grant No.
202/07/P245) (F. \v{S}.), NSF (Grant No CHE-0446555) and NIH (GM59230) (S.M.)is
gratefully acknowledged.

\newpage

\textbf{\Large Table captions}\newline
\begin{description}
\item{Table 1} $S_I$, $S_{II}$ lineshapes
shows  divergent growth  along the  $ \omega _{3}=\Omega _{eg}-\Omega _{0}$ and
 $\omega _{1}=\Omega _{eg}-\Omega _{0}$ lines. Table I shows their asymptotic form.
\end{description}

\newpage

\textbf{\Large Figure captions}\newline

\begin{description}
\item {Fig 1} Pulse configuration and time variables for a four wave mixing
experiment .

\item {Fig 2} Feynman diagrams for the third order response of a two level
chromophore with wavevector $\mathbf{k_I=-k_1+k_2+k_3}$ and $\mathbf{
k_{II}=k_1-k_2+k_3}$.

\item {Fig 3} The 8 contributions to the Green's function (Eq. (\ref{8terms}))
of the third order response. Contributions represent paths with (when the
line touch the axis) or without (when the line does not touch the axis) some
jump during each of the three time intervals $t_1$,$t_2$,$t_3$.

\item {Fig 4} Integration time variables in Eq. (\ref{allconvolutions}).

\item {Fig 5} (Color online) Linear absorption for slow $\kappa_1 \Omega_0=2$
(top panel) and fast $\kappa_1 \Omega_0=0.2$ fluctuations and different $\alpha$
as indicated. $\Omega_0 \kappa_{A}= 0.5$.

\item {Fig 6A} (Color Online) The $S_I(\omega_3,0,-\omega_1)$ (top), $S_{II}(\omega_3,0,\omega_1)$ (middle), and $S_A(\omega_3,0,\omega_1)$
(bottom) signals (Eq.(\ref{2DIRdef})) for the WTDF (Eq. (\ref{weak_anomalous}))
at $t_2=0$, for slow fluctuations $\Omega_0 \kappa_1= 2.0$, and
$\kappa_{A}/\kappa_1=0.25$, $\alpha =1.2$ (left), $1.5$ (middle), $1.8$
(right).

\item {Fig 6B} (Color Online) The same as Fig 6a but for fast fluctuations
$\Omega_0 \kappa_1= 0.2$, and $\kappa_{A}\Omega_0=0.5$, $\alpha =1.2$
(left), $1.5$ (middle), $1.8$ (right).

\item {Fig 7} (Color Online) The $S_A$ signal (Eq.(\ref{2DIRdef})) for the
WTDF (Eq. (\ref{weak_anomalous})) for (left to right) $\alpha=1.2,1.5,1.8$,
and $\kappa_{A}/\kappa_1=0.25$, $\Omega_0 \kappa_1= 2.0$, $t_2=\kappa_1$
(top), $t_2=2\kappa_1$(middle), $t_2=10\kappa_1$(bottom).

\item {Fig 8} (Color Online) Aging effects in 2D lineshapes. The $S_A$
signal (Eq.(\ref{2DIRdef})) for the nonstationary random walk model
(Eq. (\ref{mittag})) $t_2=0$, $\kappa \Omega_0=0.2$, $\alpha =0.98$ for
various initial time (from left top, to right bottom) $t_0=0\kappa$, $10\kappa$,
 $10^2\kappa$, $10^3\kappa$, $10^4\kappa$, $10^5\kappa$.

\item{Fig 9} (Color online) Time-dependent rates of aging random walk Eq. (\ref{defrates})
for $\alpha= 0.3$ (solid), $0.5$ (dashed),$0.7$ (short-dashed),and
$0.98$ (dotted line).

\item{Fig 10} (Color online) Aging 2D Markovian lineshape (Eq. (\ref{ml})) for various initial time $t_0=10^{-3}$,
$1,5,10,20,$ and $100$ . Master equation for probability
densities correspond to the random walk of Fig. 8
\end{description}

\newpage
\begin{tabular}{|c|c|c|}
\hline
fixed & $\Delta\omega_3$ & $\Delta \omega_1$ \\ \hline
varied & $\Delta \omega_1$ & $\Delta \omega_3$ \\ \hline
$S_I(\omega_3,-\omega_1)\sim \frac{\mu^4}{\hbar^3}\sin{\left[\pi(2-\alpha)/2
\right]} \times $ & sgn$(\Delta\omega_1)\frac{|\Delta \omega_1|^{\alpha-2}}{
\Delta \omega_3}$ & - sgn$(\Delta\omega_3) \frac{|\Delta \omega_3|^{\alpha-2}
}{\Delta \omega_1} $ \\ \hline
$S_{II}(\omega_3,\omega_1)\sim \frac{\mu^4}{\hbar^3} \sin{\left[
\pi(2-\alpha)/2\right]} \times$ & - sgn$(\Delta\omega_1) \frac{|\Delta
\omega_1|^{\alpha-2}}{\Delta \omega_3}$ & sgn$(\Delta\omega_3)\frac{|\Delta
\omega_3|^{\alpha-2}}{\Delta \omega_1}$ \\ \hline
\end{tabular}
\newline
Table I
\newpage
\begin{center}
\scalebox{0.82}[.82]{\includegraphics{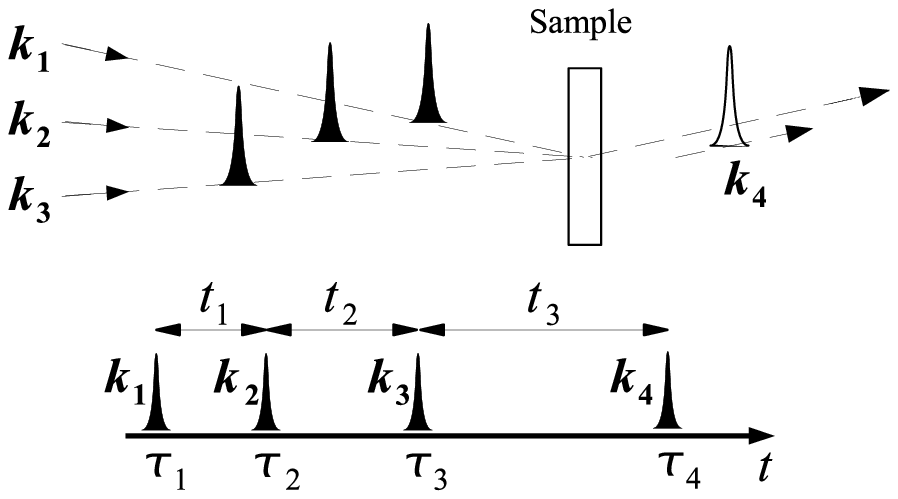}}
\end{center}
$\quad \quad \quad  \quad \quad\quad \quad \quad \quad \quad$   
{\large \bf Fig 1}

\newpage
\begin{center}
\scalebox{0.60}[0.60]{\includegraphics{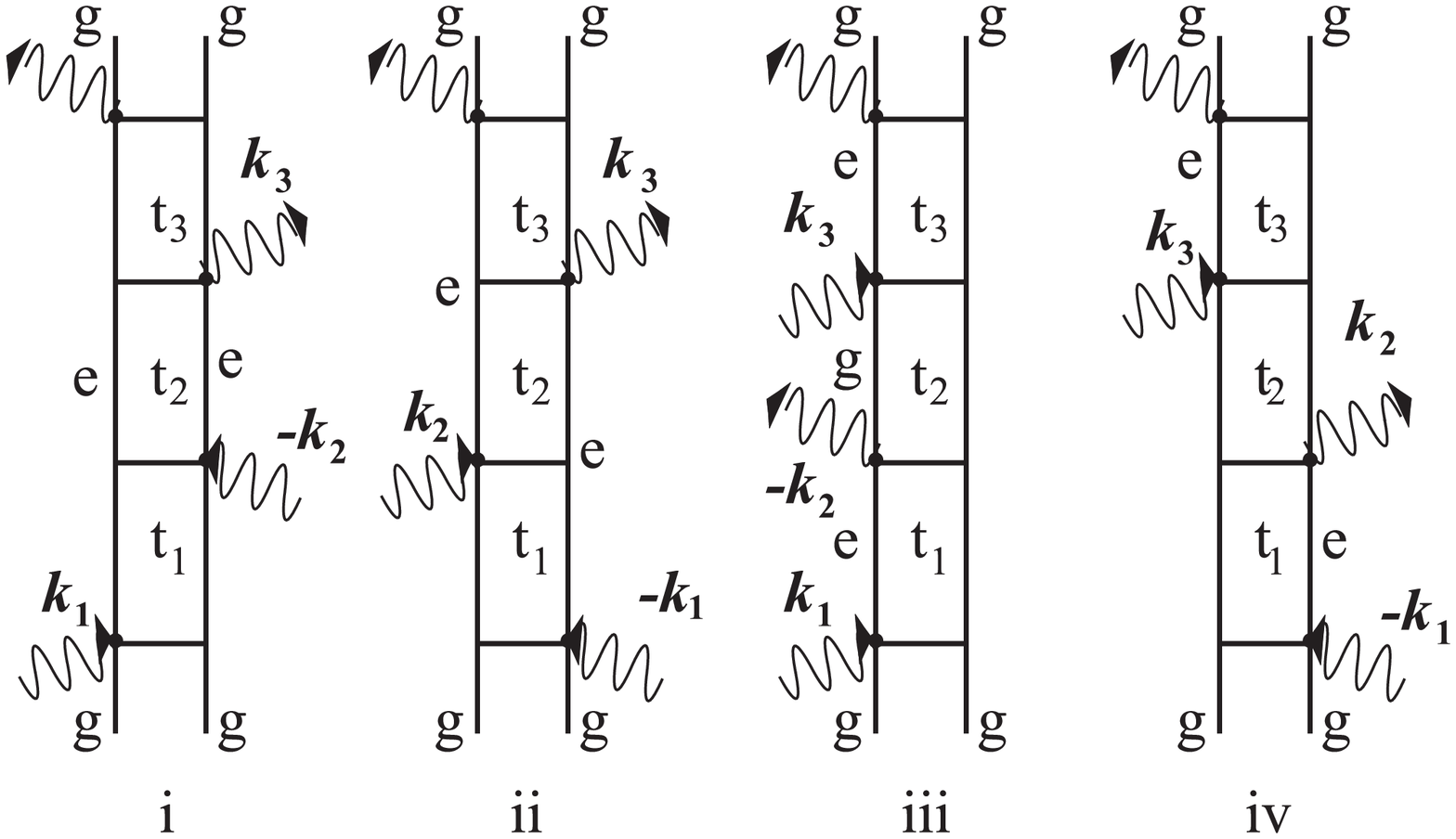}}
\end{center}
$\quad \quad \quad  \quad \quad\quad \quad \quad \quad \quad$   {\large \bf Fig 2}

\newpage
\begin{center}
\scalebox{0.60}[0.60]{\includegraphics{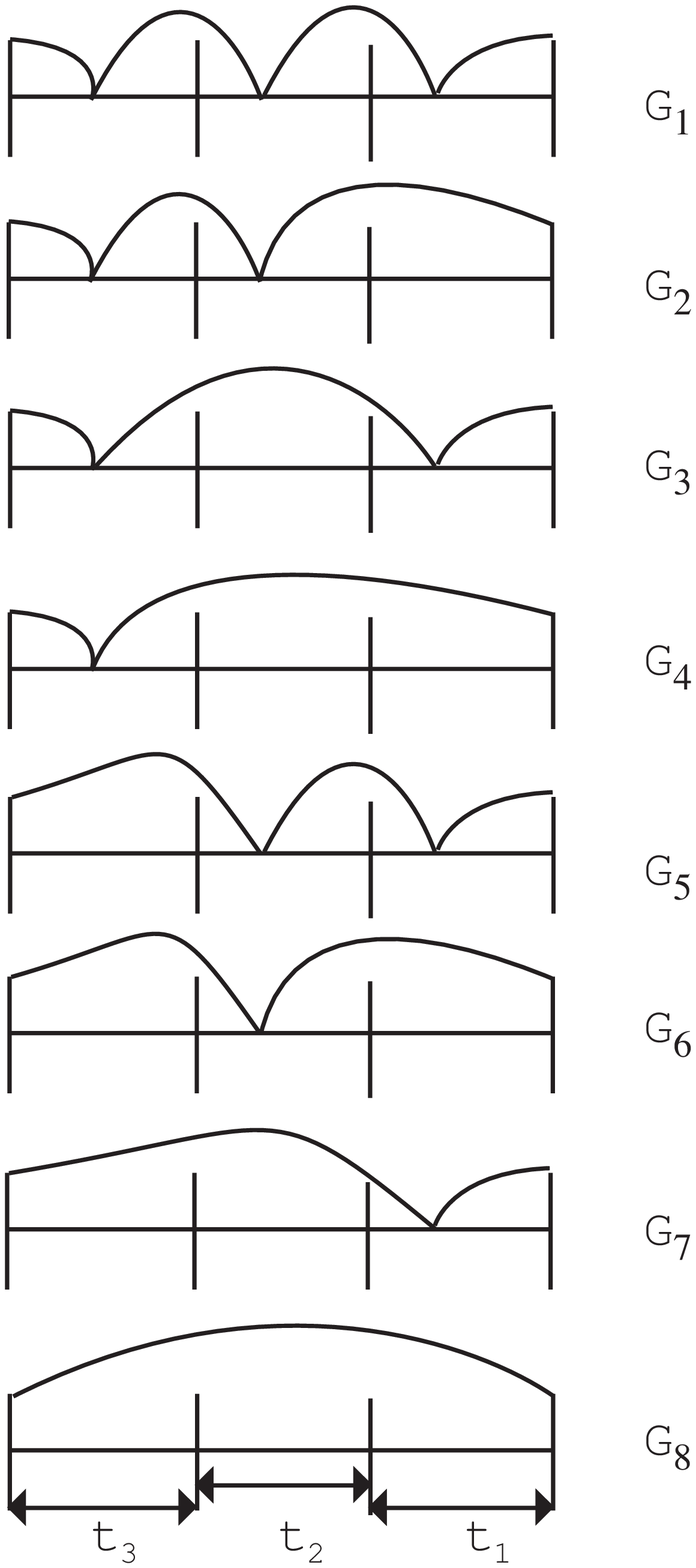}}
\end{center}
$\quad \quad \quad  \quad \quad\quad \quad \quad \quad \quad$   {\large \bf Fig 3}

\newpage
\begin{center}
\scalebox{0.60}[0.60]{\includegraphics{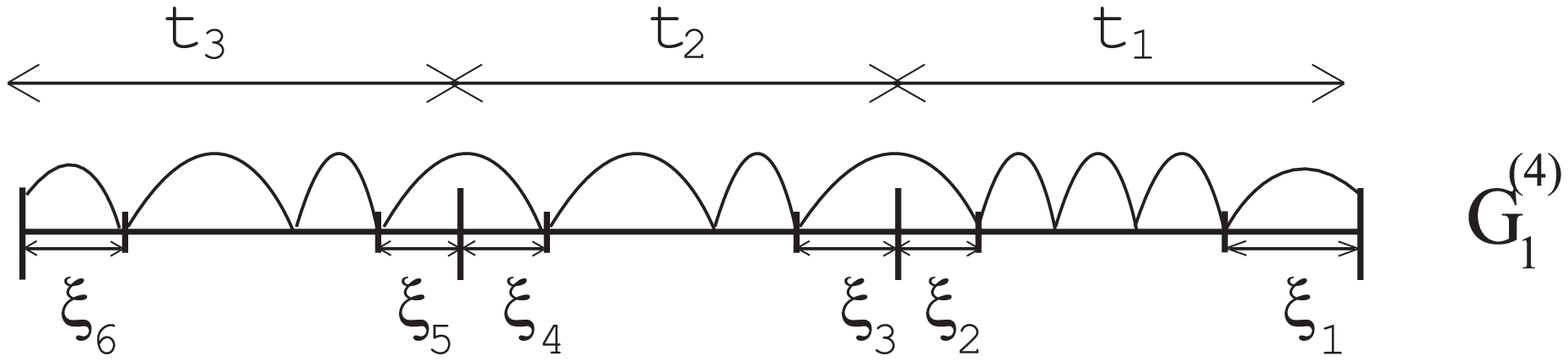}}
\end{center}
 {\large \bf Fig 4}
\newpage

\begin{center}
\scalebox{1.2}[1.2]{\rotatebox{-90}{\includegraphics{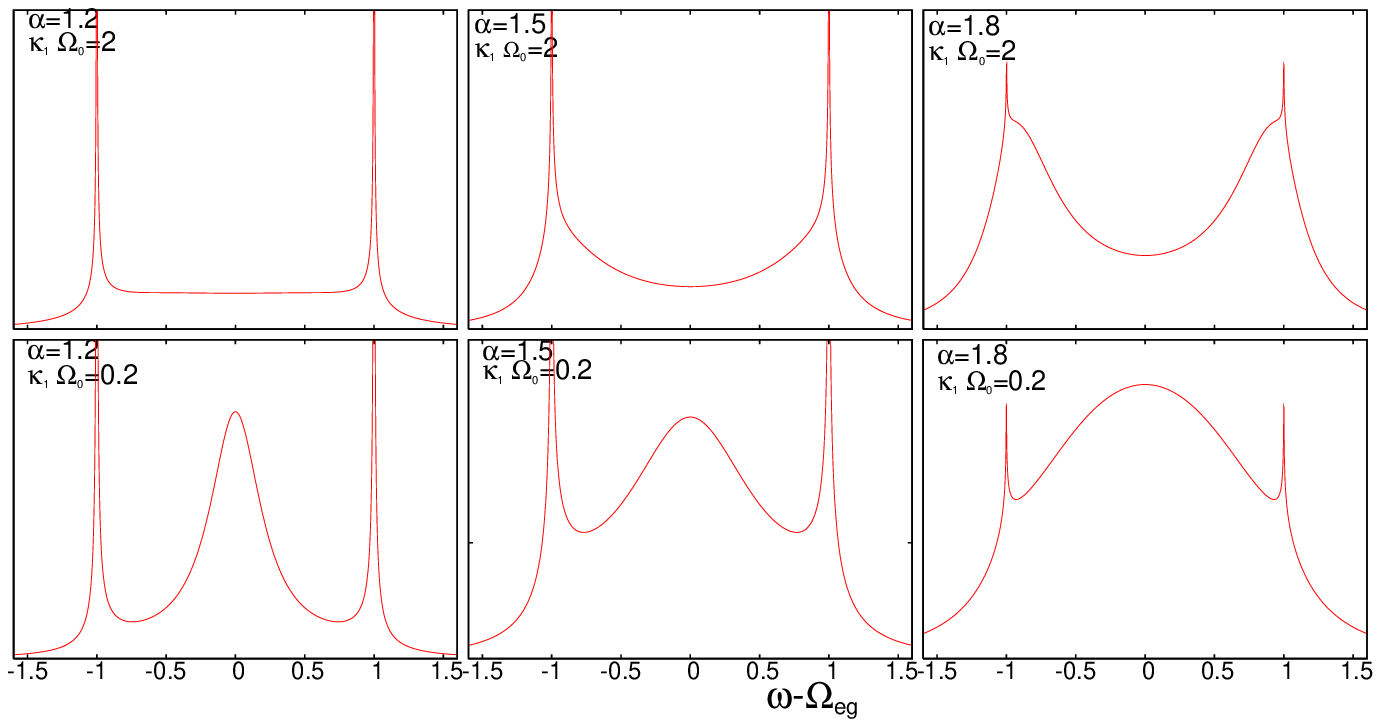}}}
\end{center}

{\large \bf Fig 5}

\newpage
\begin{center}
\scalebox{1.0}[1.0]{\rotatebox{0}{\includegraphics{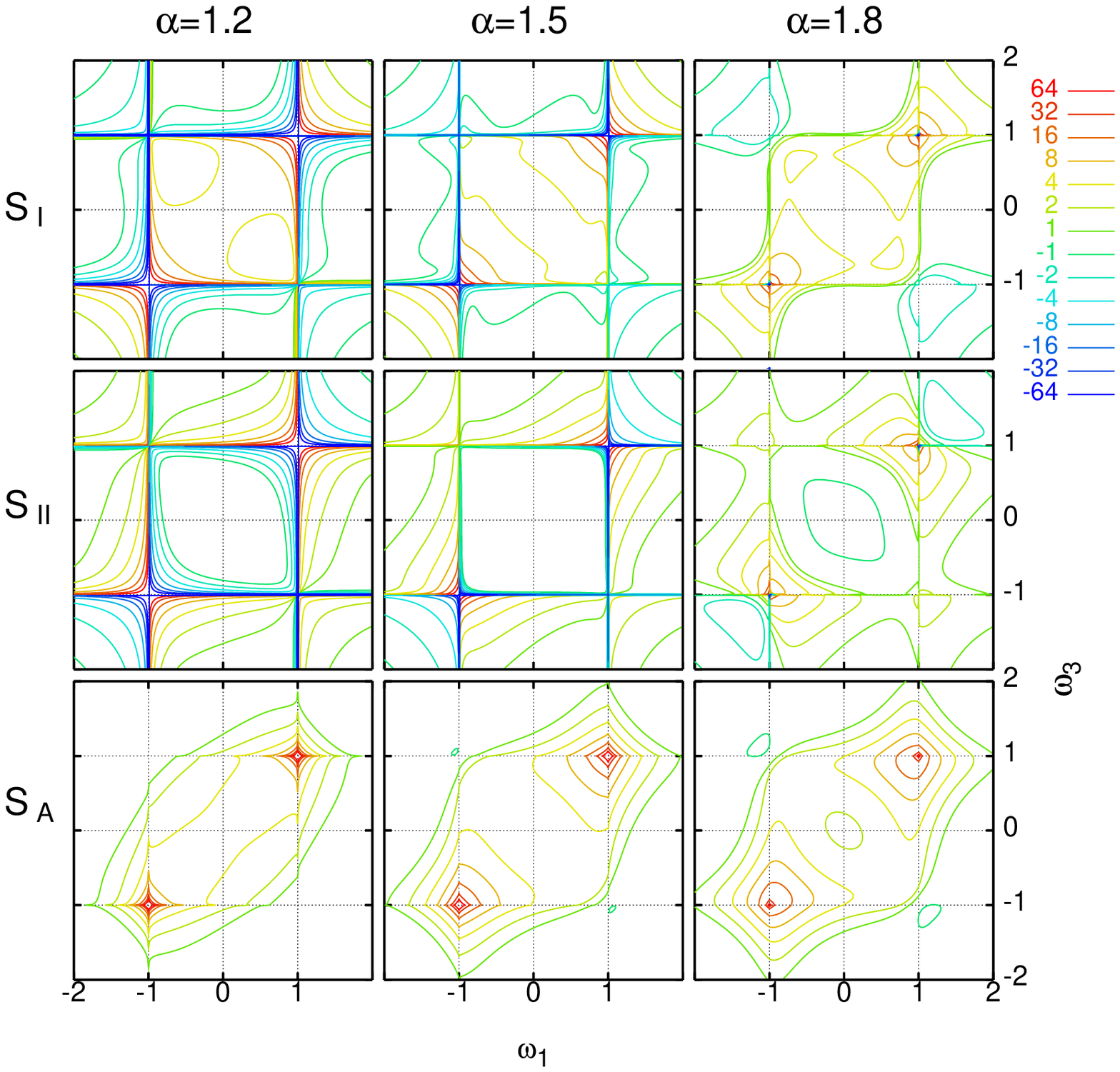}}}
\end{center}
 {\large \bf Fig 6A}
\begin{center}
\scalebox{1.6}[1.6]{\rotatebox{-90}{\includegraphics{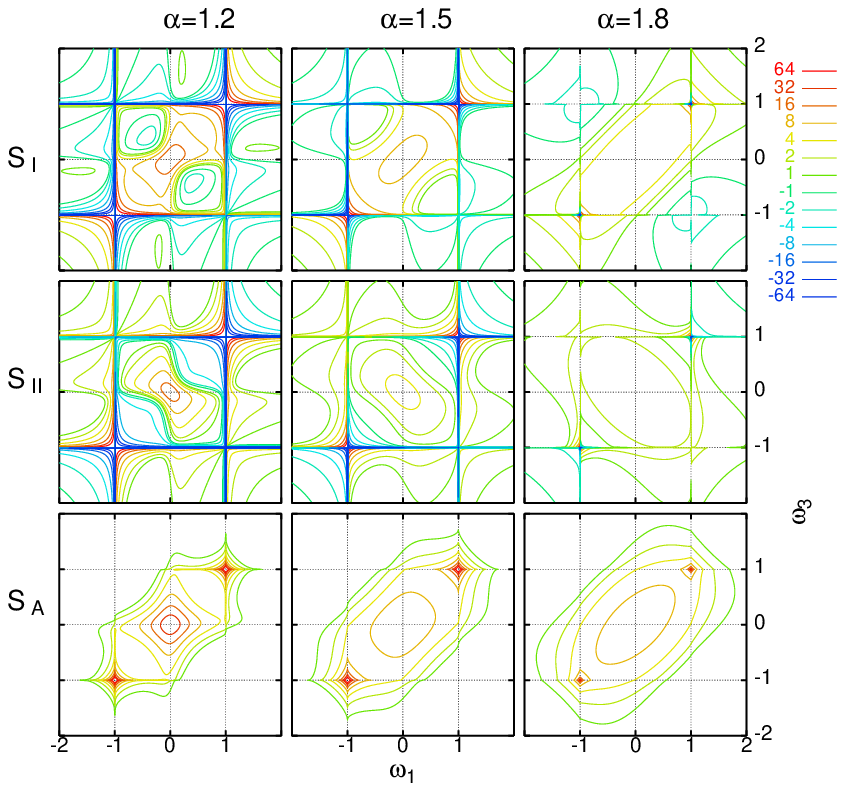}}}
{\large \bf Fig 6B}
\end{center}
\newpage
\begin{center}
\scalebox{1.60}[1.60]{\rotatebox{-90}{\includegraphics{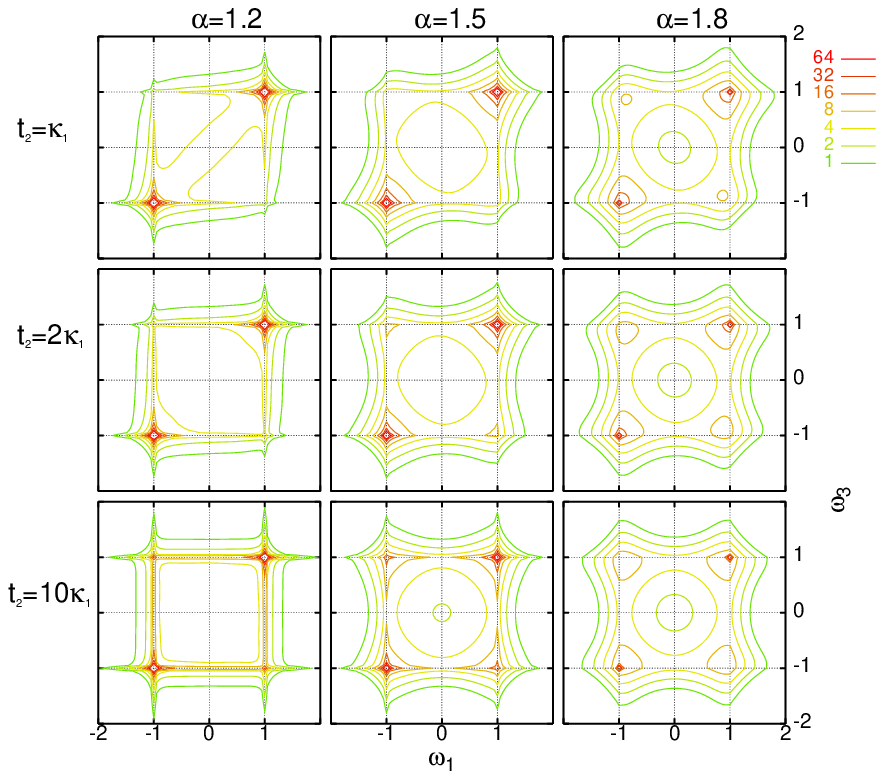}}}
\end{center}
 {\large \bf Fig 7}
\newpage
\vspace{150mm}
\begin{center}
\vspace{150mm}
\scalebox{0.83}[0.83]{\includegraphics{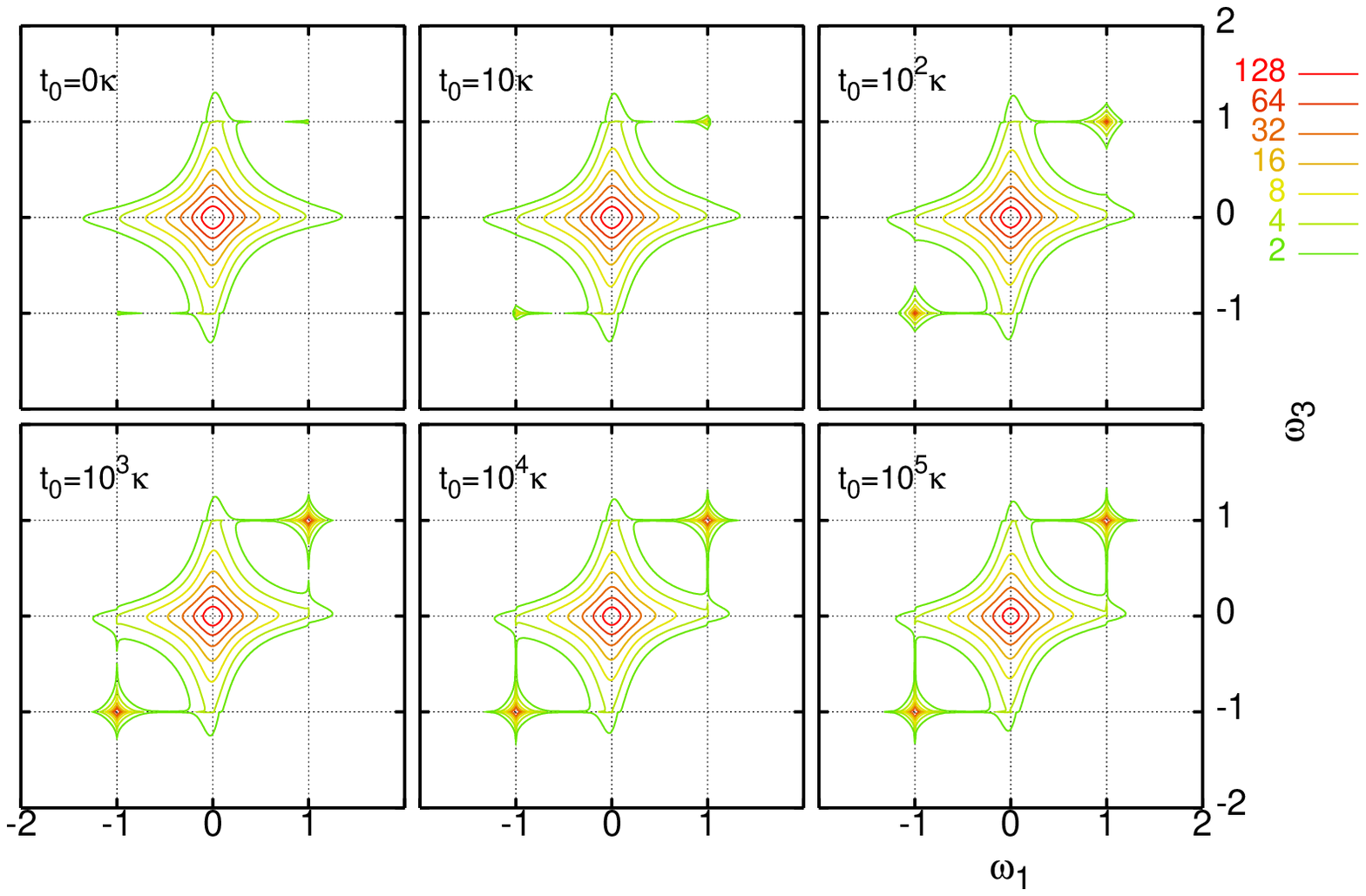}}
\end{center}
 {\large \bf Fig 8}
\newpage
\begin{center}
\scalebox{0.50}[0.50]{\rotatebox{-90}{\includegraphics{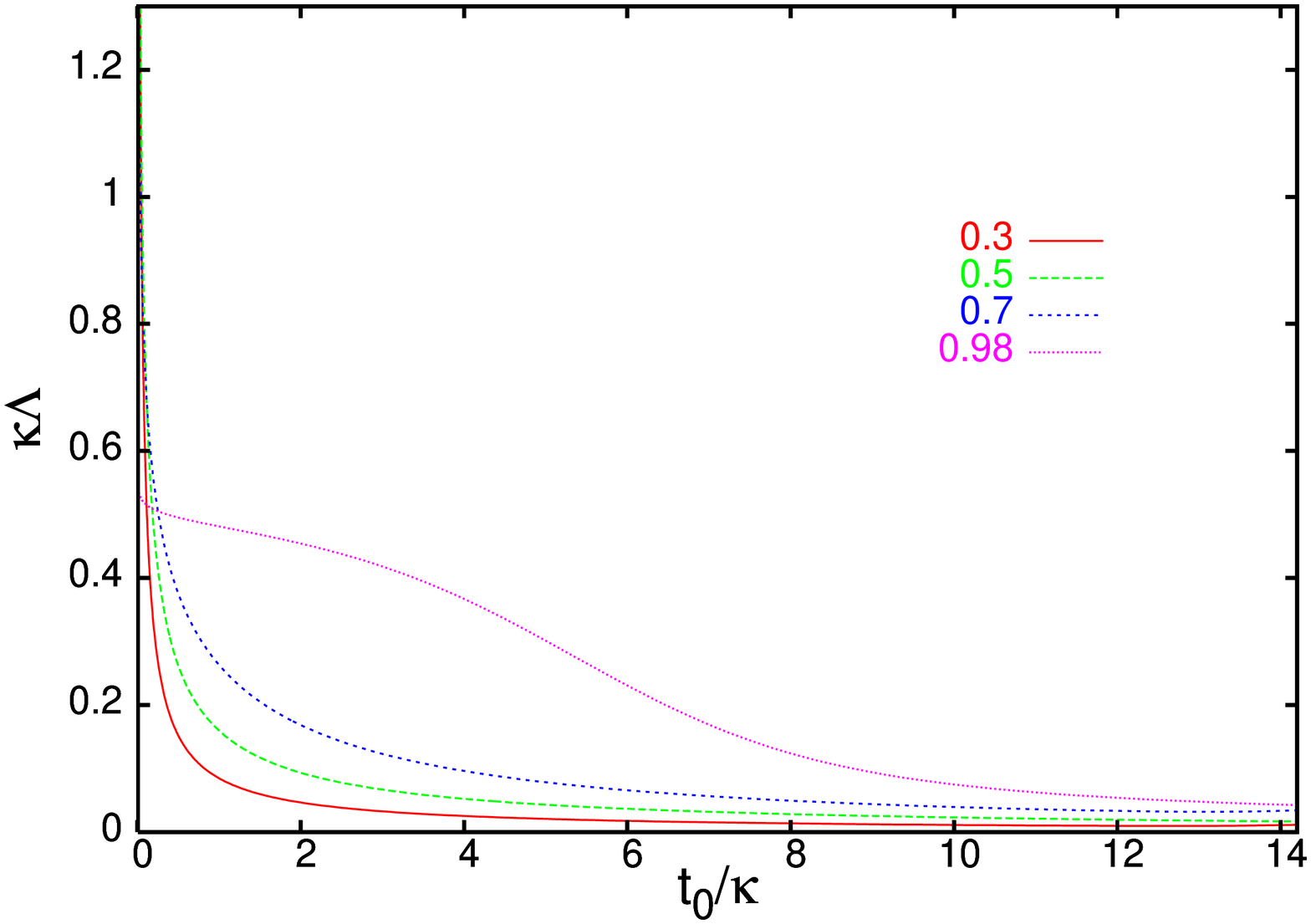}}}
\end{center}
{\large \bf Fig 9}
\newpage
\begin{center}
\scalebox{0.50}[0.50]{\rotatebox{-90}{\includegraphics{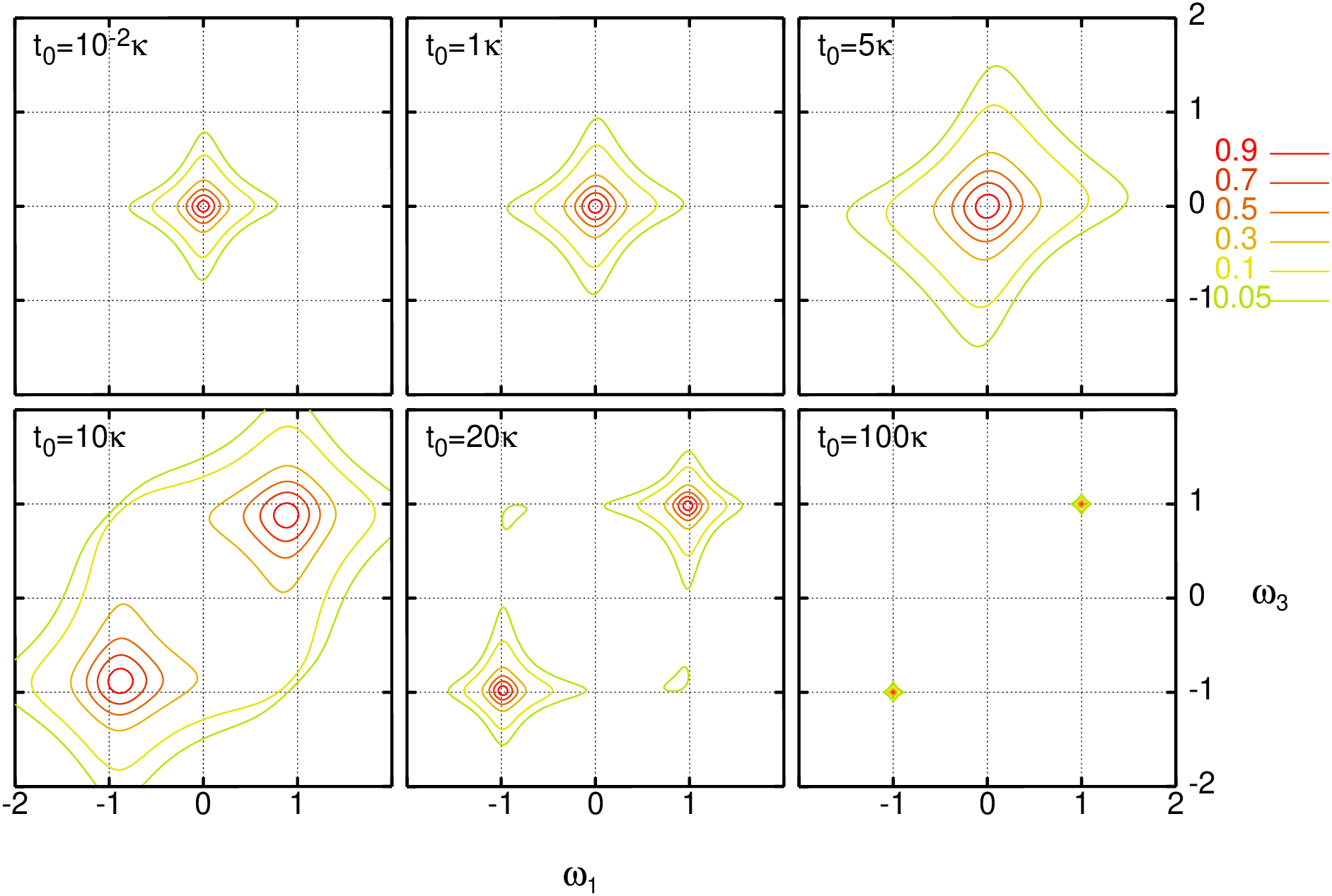}}}
\end{center}
\vspace{50mm}
{\large \bf Fig 10}


\begin{thebibliography}{99}
\bibitem{note3} This condition is satisfied for any TSJ random walk (even
asymmetric $[\Psi]_{ab} \neq [\Psi]_{ba}$). Further important class of such
random walks are separable WTDF $[\Psi]_{ji}=T_{ji}\psi(t)$, provided that
$T $ is detail balanced.
\bibitem{note1} When the system's evolution (and equilibrium) depends on a
state of chromophore, which goes beyond the present stochastic model, there
will be time directionality of contributions evolving in $ee$ state during
$t_2$ $R_{i}$, $R_{iii}$. Eq. (\ref{sym_diagonal}) is thus broken. For the
same reasons in the long $t_2$ limit Eq. (\ref{s1as}) and Eq. (\ref{s2as})
are modified to $R_{iii}(t_3,t_2,t_1)= (i/\hbar)K_e(t_3) K(t_1)$, and
$R_{iv}(t_3,t_2,t_1)=K_e(t_3) K^*(t_1)$ where $K_e$ represents linear
response of the excited state equilibrium at the beginning of $t_3$
interval. Eq. (\ref{factorization})) should be then replaced by
\[
\hbar S_A(\omega_3,t_2 \rightarrow \infty,\omega_1)= 2W_A(\omega_1) \left[
W_A(\omega_3)+W_E(\omega_3)\right].
\]
where $W_E(\omega)\equiv Re \int_0^{\infty} K_e(t)\exp{\left[
i(\omega-\Omega_{eg})t\right]}d t$ is the emission lineshape.
\end{thebibliography}
\end{document}